\documentclass[apj]{emulateapj}
\usepackage{hyperref}
\usepackage{float}

\slugcomment{as of \today}
\shorttitle{Diffusion in 47 Tuc}
\shorttitle{Heyl et al.}

\begin{document}

\title{ A Measurement of Diffusion in 47 Tucanae}
\author{Jeremy~Heyl\altaffilmark{1}, Harvey~B.~Richer\altaffilmark{1}, 
Elisa Antolini\altaffilmark{2}, Ryan Goldsbury\altaffilmark{1},  Jason Kalirai\altaffilmark{3,4}, Javiera Parada\altaffilmark{1}, Pier-Emmanuel Tremblay\altaffilmark{3}} 
\altaffiltext{1}{Department of Physics \& Astronomy, University of
  British Columbia, Vancouver, BC, Canada V6T 1Z1;
  heyl@phas.ubc.ca, richer@astro.ubc.ca  } 
\altaffiltext{2}{Dipartimento di Fisica e Geologia,
  Universit\`a degli Studi di Perugia, I-06123 Perugia, Italia}
\altaffiltext{3}{Space Telescope Science Institute,Baltimore MD  21218}
\altaffiltext{4}{Center for Astrophysical Sciences, Johns Hopkins University, Baltimore MD, 21218; jkalirai@stsci.edu }
 
\begin{abstract}

Using images from the Hubble Space Telescope Wide-Field Camera 3, we
measure the rate of diffusion of stars through the core of the
globular cluster 47~Tucanae using a sample of young white dwarfs
identified in these observations.  This is the first direct
measurement of diffusion due to gravitational relaxation.  We find
that the diffusion rate $\kappa\approx 10-13$ arcsecond$^2$~Myr$^{-1}$
is consistent with theoretical estimates of the relaxation time in the
core of 47 Tucanae of about 70~Myr.  
\end{abstract}

\keywords{globular clusters: individual (47 Tuc) --- stars: Population II, Hertzsprung-Russell and C-M diagrams, kinematics and dynamics}

%
\section{Introduction}
\label{sec:intro}

Globular clusters have long provided an amazing laboratory for stellar
evolution and gravitational dynamics, and the nearby rich cluster, 47
Tucanae, has long been a focus of such investigations.  The key point
of this investigation is an interplay between these two processes.  In
particular in the core of 47 Tucanae, the timescale for stellar
evolution and the timescale for dynamical relaxation are similar.  The
relaxation time in the core of 47 Tuc is about 70~Myr
\citep{1996AJ....112.1487H}.  Meanwhile over a span of about 150~Myr
the most massive stars in 47~Tucanae evolve from a red giant star with
a luminosity of 2,000 times that of the Sun to a white dwarf with a
luminosity less than a tenth that of the Sun.  Meanwhile the star
loses about forty percent of its mass, going from 0.9 to 0.53 solar
masses.  It is these young white dwarfs that are the focus of this
paper.

Although the core of 47~Tuc has been the focus of numerous previous
investigations
\citep[e.g][]{2006ApJS..166..249M,2008ApJ...683.1006K,2009AJ....138.1455B},
this is the first paper that combines the near ultraviolet filters of
the Hubble Space Telescope (HST) with a mosaic that covers the entire
core of the cluster.  Probing the core of the cluster in the
ultraviolet is advantageous in several ways.  First the young white
dwarfs are approximately as bright as the upper main sequence, giant
and horizontal branch stars at 225~nm, so they are easy to find.  In
fact the brightest white dwarfs are among the brightest stars in the
cluster and are as bright as the blue stragglers.  Second, the
point-spread function of HST is more
concentrated in the ultraviolet helping with confusion in the dense
starfield that is the core of 47~Tuc.

In spite of these advantages, for all but the brightest stars, our
dataset suffers from incompleteness which presents some unique
challenges.  We reliably characterize the incompleteness as a function
of position and flux in the two bands of interest F225W and F336W
throughout the colour-magnitude diagram and especially along the
white-dwarf cooling sequence through the injection and recovery of
about $10^8$ artificial stars into the images. How we measure the
completeness is described in detail in \S\ref{sec:artif-star-tests}.
The young white dwarfs typically have a mass forty percent less than
their progenitors, so they are born with less kinetic energy than
their neighbors, and two-body interactions will typically increase the
kinetic energy of young white dwarfs over time and change their
spatial distribution.  We introduce a simple model for the diffusion
of the young white dwarfs through the core of the cluster
(\S\ref{sec:difuss-lumin-evol}).  To make the most of this unique
dataset, we have to include stars in our sample whose completeness
rate is well below fifty percent.  We have developed and tested
statistical techniques to characterize the observational distribution
of young white dwarfs in flux and space to understand their motion
through the cluster and their cooling
(\S\S\ref{sec:likelihood-function}-\ref{sec:lferr}) in the face of
these potentially strong observational biases.  Although these
techniques are well known especially in gamma-ray astronomy, they have
never been applied to stellar populations in this way, so
\S\ref{sec:monte-carlo-simul} presents a series of Monte Carlo
simulations to assess the potential biases of these techniques and
verify that these techniques are indeed unbiased in the face of
substantial incompleteness within the statistical uncertainties.  To
establish the time over which the white dwarfs dim we use a stellar
evolution model outlined in \S\ref{sec:cooling-model}.
\S\ref{sec:results} describes the best-fitting models for the density
and flux evolution of the white dwarfs.
\S\ref{sec:two-body-relaxation} looks at the dynamic consequences of
these results. \S\ref{sec:conclusions} outlines future directions both
theoretical and observational and the broader conclusions of this
work.

\section{Observations}
\label{sec:observations}

\begin{figure*}
\includegraphics[width=\textwidth,clip,trim=80px 30px 100px 20px]{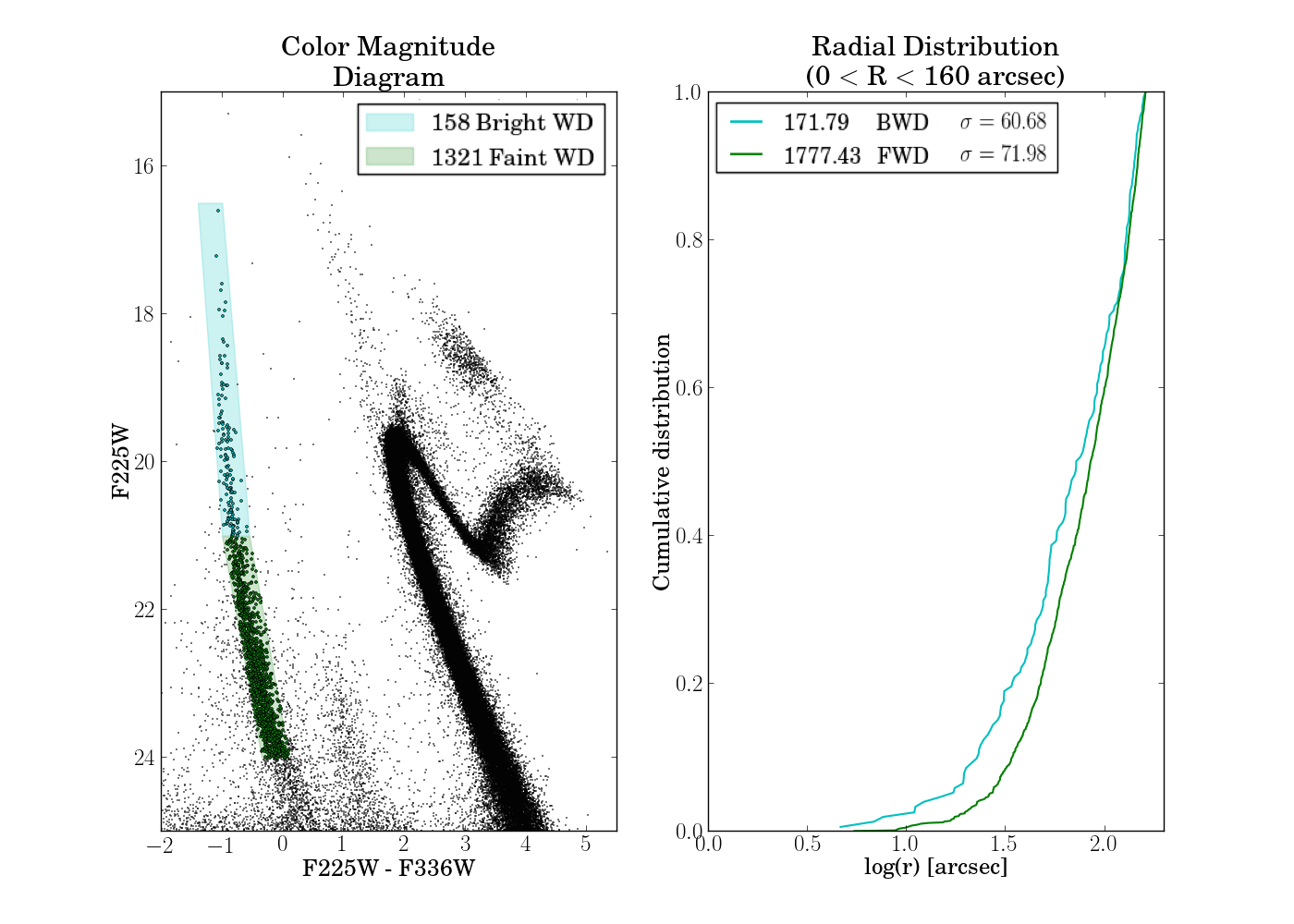}
\caption{Left: Color-magnitude diagram of the core of 47 Tucanae in
  the WFC3 filters F225W and F336W. Right: The radial distribution of
  young and old white-dwarf stars as highlighted in the color-magnitude
  diagram (completeness corrected numbers appear right-hand panel).}
\label{fig:CMD}
\end{figure*}

A set of observations with the Advanced Camera for Surveys
\citep[ACS,][]{1998SPIE.3356..234F} and the Wide Field Camera 3
\citep[WFC3,][]{2012SPIE.8442E..1VM} on the Hubble Space Telescope
(HST) of the core of the globular cluster 47 Tucanae over one year
provides a sensitive probe of the stellar populations in the core of
this globular cluster (Cycle 12 GO-12971, PI: Richer), especially the
young white dwarfs.  Here we will focus on the observations with WFC3
in the UV filters, F225W and F336W.  The observations were performed
over ten epochs from November 2012 to September 2013. Each of the
exposures in F225W was 1080 seconds, and the exposures in F336W were
slightly longer at 1205 seconds.  Each of the overlapping WFC3 images was
registered onto the same reference frame and drizzled to form a single
image in each band from which stars were detected and characterized,
resulting in the color-magnitude diagram depicted in
Fig.~\ref{fig:CMD}.

What is immediately striking in Fig.~\ref{fig:CMD} is that the
distribution of young white dwarfs with a median age of 6~Myr is
significantly more centrally concentrated than that of the older white
dwarfs that have a median age of 127~Myr.  The white-dwarf
distribution appears to become more radially diffuse with increasing
age, a signature of relaxation. One concern is immediately apparent.
The numbers of observed stars are given in legend of the left panel
and the numbers of stars in the completeness corrected samples are
given in the legend of the right panel.  The sample of older white
dwarfs is only about seventy-five percent complete on average.
Furthermore, one would expect the completeness of these faint stars to
be lower near the centre of the cluster, so if the completeness is not
accounted for correctly, one could naturally conclude that the white
dwarfs are diffusing when they are not in reality.  In principle we
would like to divide this sample of over 1,300 stars into subsamples
some of which will have even smaller completeness rates.  How can we
be sure that our analysis techniques are up to the task of measuring
this diffusion accurately in the face of completeness rates as low as
twenty percent that vary dramatically with distance from the center of
the cluster?

In the following section (\S~\ref{sec:analysis}) we will characterize
the completeness rate through artificial star tests, develop and test
statistical techniques to measure the diffusion of white dwarfs in
47~Tucanae without binning the stars at all, thus preserving the
maximal information content of these data.  We will test these new
algorithms on mock data sets that include both the completeness rate
and flux error distribution of our sample to verify that they robustly
determine the diffusion and flux evolution of the white dwarfs.  The
subsequent section (\S~\ref{sec:results}) explores results of these
techniques on the dataset depicted in Fig.~\ref{fig:CMD}.

\section{Analysis}
\label{sec:analysis}

\subsection{Artificial Star Tests}
\label{sec:artif-star-tests}

\begin{figure}
\includegraphics[width=\columnwidth,clip,trim=1.6in 0.5in 0.95in 0.6in]{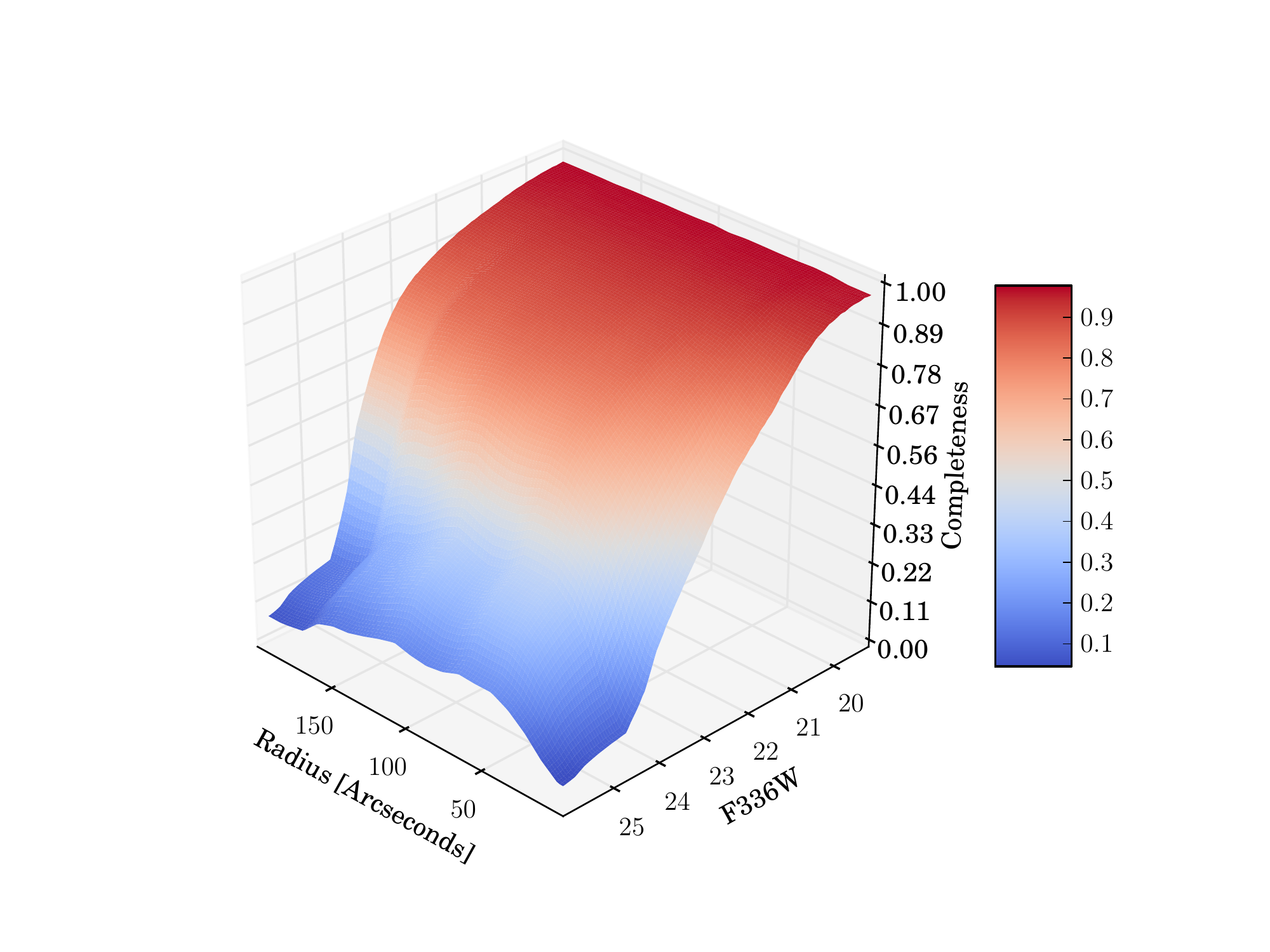}
\caption{The completeness rate as function of radius from the center of 47 Tuc and
the magnitude of the artificial star.}
\label{fig:comp336}
\end{figure}

We inserted $\sim 10^8$ artificial stars into the WFC3 images both in
F225W and F336W over the full range of observed magnitudes in both
bands and a range of distances from the center of the cluster.  To
determine the completeness rate for the white dwarfs that we have
observed, we inserted artificial stars whose F225W and F336W
magnitudes lie along the observed white-dwarf track in the CMD.  The
rate of recovering a star along the white-dwarf track of a given input
magnitude in F336W at a given radius is the completeness rate and is
depicted in Fig.~\ref{fig:comp336}. If an artificial star along the
white-dwarf track is detected in F336W, it is always detected in F225W
as well.  The completeness rate is both a strong function of radius
and magnitude and is significantly different from unity except for the
brightest stars, so accounting for completeness robustly is crucial in
the subsequent analysis.  The radial bins are 100 pixels in width and
the magnitude bins are 0.1358 wide.

The magnitudes of the recovered stars give the error distribution as a
function of the input magnitude and position of the star in the field.
Furthermore, these distributions are not typically normal and often
asymmetric as well.  For the analysis in \S\ref{sec:magnitude-errors}
we use the cumulative distribution of magnitude errors as a function
of position and input magnitude which we obtain by sorting the output
magnitudes in a given bin and spline to obtain the cumulative
distribution in the form of the values of the errors from the first to
the ninety-ninth percentile.  In the analysis the completeness rate is
interpolated over the two dimensions of radius and magnitude with a
third-degree spline, and the error distributions are interpolated
linearly over the three dimensions (radius, magnitude and percentile).

\subsection{Diffusion and Luminosity Evolution}
\label{sec:difuss-lumin-evol}

Sometime during the late evolution of a turn-off star in 47~Tuc, the
star loses about forty percent of its mass, going from a main-sequence
star of ninety percent of a solar mass to a white dwarf of fifty-three
percent of a solar mass
\citep{1988ARA&A..26..199R,1996ApJ...465L..23R,2004A&A...420..515M,2009ApJ...705..408K}.
These newborn white dwarfs will have the typical velocities of their
more massive progenitors, so as they interact gravitationally with
other stars, their velocities will increase through two-body
relaxation, bringing their kinetic energies into equipartition \citep[e.g.][]{Spit87}.
Because the gravitational interaction is long range and the distance
between the stars is small compared to the size of the cluster, the
change in velocity will be dominated by distant interactions and small
random velocity jumps, {\em i.e.} the Coulomb logarithm is large,
$\sim\ln N$ where $N$ is the number of stars in the core of the
cluster, $\sim 10^{5-6}$.  These small jumps in velocity can be
modeled as a random walk in velocity so the square of the
velocity increases linearly in time and the relaxation time can be
defined as $t_r = \left [d\ln v^2/dt \right]^{-1}$.
In the center of the
cluster, the density of stars is approximately constant, so the
gravitational potential has the approximate form
\begin{equation}
\phi = \frac{2}{3} \pi G \rho_c r^2.
\label{eq:42}
\end{equation}
By the virial theorem the mean kinetic energy of the white dwarfs will
equal the mean potential energy, 
\begin{equation}
\frac{1}{2} \langle v^2 \rangle = \frac{2}{3} \pi G \rho_c \langle r^2 \rangle
\label{eq:43}
\end{equation}
so the square of the distance of the
white dwarfs from the center of the cluster will also increase
linearly with time as a random walk; therefore, let us suppose that
newly born white dwarfs diffuse outward through the cluster following
the diffusion equation
\begin{equation}
\frac{\partial \rho(r,t)}{\partial t} = \kappa \nabla^2 \rho(r,t)
\label{eq:33}
\end{equation}
where $\kappa$ can be related to the relaxation time as
$\kappa=r_c^2/t_r$ because $d\ln r^2/dt=d\ln v^2/dt$ from Eq.~\ref{eq:43}.
This diffusion equation yields the Green's function
\begin{equation}
u(r,t) = \frac{1}{8\left (\pi \kappa t\right)^{3/2}} e^{-r^2/(4 \kappa t)}
\label{eq:34}
\end{equation}
if $\kappa$ is independent of time and position.  This gives
a cumulative distribution in projected radius
\begin{equation}
C(<R) = 1 - e^{-R^2/(4 \kappa t)}.
\label{eq:35}
\end{equation}
The Green's function at $t=0$ is a delta function centered on the
center of the cluster.  On the other hand, if the initial distribution
is a Gaussian centered on the center of the cluster the density of
white-dwarf stars near the center of the cluster is a function of age,
$t$, and projected radius, $R$, of the form
\begin{equation}
\rho = \rho(R,t) = \frac{2 R}{4 \kappa ( t + t_0 )} \exp \left[ - \frac{R^2}{4 \kappa (t + t_0)}
\right ]
\label{eq:1}
\end{equation}
where the density distribution is normalized as
\begin{equation}
\int_0^\infty \rho(R,t) dR = 1.
  \label{eq:2}
\end{equation}
The dispersion of the Gaussian at $t=0$ is simply given by $2\sigma^2=4 \kappa t_0$.
Because the diffusion equation is linear, a sum of several
Gaussians with the same value of $\kappa$ but different normalizations
and values of $t_0$ will also be a solution.

Of course we don't directly observe the ages of the white dwarfs.
Rather we observe their fluxes or apparent magnitudes.  The cooling
curve of the white dwarfs is a relationship between time and the
apparent magnitude from the white dwarfs $t(m)$, so the number of
white dwarfs that we expect to observe at a given flux and radius is
given by
\begin{equation}
f(R,m) = \dot N \rho(R,t(m)) \frac{\partial t}{\partial m} C(R,m)
\label{eq:41}
\end{equation}
where $\dot N$ is the birthrate of the white dwarfs (assumed to be
constant over the range of ages of the young white dwarfs, i.e. the
past 200 Myr) and $C(R,m)$ is the completeness as a function of radius
and flux. To this point flux errors have been neglected.

\subsection{Cooling Models}
\label{sec:cooling-model}

To construct the various cooling models here, {\em i.e.} $t(m)$ from
Eq.~\ref{eq:41}, we used MESA (Modules for Experiments in Stellar
Astrophysics; \citealt{2011ApJS..192....3P}) to perform simulations of
stellar evolution starting with a pre-main-sequence model of 0.9 solar
masses and a metallicity of $Z=3.3\times 10^{-3}$ appropriate for the
cluster 47 Tucanae.  This is slightly larger than the value for the
turnoff mass found by \cite{2010AJ....139..329T} for the eclipsing
binary V69 in 47~Tucanae that is composed of a upper-main sequence
star of about 0.86 solar masses and a subgiant of 0.88 solar masses.
Because we are interested in the stars that have become young white
dwarfs just recently, the initial masses of these stars should be
slightly larger than the turnoff mass today.  We have explicitly
assumed that the progenitors of the white dawrfs are a uniform
population.  Although there is evidence of modest variation in the
chemical abundances in 47~Tuc \citep[e.g.][]{2012ApJ...744...58M}, the
white-dwarf cooling sequence, at least at larger radii, appears
uniform \citep{2013ApJ...778..104R}.  However, from Fig.~\ref{fig:CMD}
it is apparent that the core of 47~Tuc has a substantial population of
blue stragglers that will evolve to become more massive white dwarfs.
Our sample has about 160 blue stragglers, and if we estimate the
duration of the main sequence for a blue straggler to be $1-2$~Gyr
\citep[e.g][]{2009ApJ...692.1411S}, we obtain a birth-rate of
blue-straggler white dwarfs of about 0.1~Myr$^{-1}$. The number of
giants in our field indicates a birth rate of about eight white dwarfs
per million years \citep[see][for further details]{coolpap}, so the estimated
contamination of the white-dwarf cooling sequence is modest at about
one percent.

Specifically, we used SVN revision 5456 of MESA and started with the
model {\tt 1M\_pre\_ms\_to\_wd} in the test suite.  We changed the
parameters {\tt initial\_mass} and {\tt initial\_z} of the star and
adjusted the parameter {\tt log\_L\_lower\_limit} to $-6$ so the
simulation would run well into the white dwarf cooling regime.  We
also reduced the two values of the wind $\eta$ to 0.46 (from the
default of 0.7) to yield a 0.53 solar mass white dwarf from the 0.9
solar mass progenitor.  Interestingly \citet{2012MNRAS.419.2077M}
argue from Kepler asteroseismic measurements of the stars in the
metal-rich open cluster NGC~6971 that such values of $\eta$ are needed
to account for the mass loss between the red giant and red clump
phases of stars in this metal-rich cluster.

We defined the time of birth of the white dwarf to coincide with the
peak luminosity of the model at the tip of the asymptotic giant branch
about 10.9~Gyr after the start of the simulation.  This is in
agreement with the best age of the cluster determined from
main-sequence stars of $11.25\pm 0.21$ (random) $\pm 0.85$
(systematic) Gyr \citet{2010AJ....139..329T}.  This age agrees with
that derived by \citet{2013Natur.500...51H} from white-dwarf cooling
($9.9\pm 0.7$~Gyr at 95\% confidence).  We choose this definition of
the birth so that each observed white dwarf will have a star of
similar luminosity in the cooling model. At this point in the
evolution we have outputs from the MESA evolution every 100 years or
so; therefore, the cooling curve is well sampled throughout.  At each
output time we have the value of the luminosity, radius, effective
temperature and mass of the star.  With these values we interpolate
the spectral models of \citet{2011ApJ...730..128T} in surface gravity
and effective temperature and then scale the result to the radius of
the model star.  We use a true distance modulus of 13.23
\citep{2010AJ....139..329T} and a reddening of $E(B-V)=0.04$
\citep{2007A&A...476..243S} to determine the model fluxes in the WFC3
band F336W.  We used the standard extinction curve of
\citet{1999PASP..111...63F} with $A_V/E(B-V) = 3.1$. We have
purposefully used a distance and reddening determined from main
sequence stars to avoid a potential circularity in using the white
dwarf models themselves to fix the distance.
\citet{2012AJ....143...50W} inferred a slightly larger true distance
modulus of $13.36\pm0.02\pm0.06$ from the white-dwarf spectral energy
distributions.

The brightest white dwarf in our sample has $\mathrm{F336W}=14.92$.
According to the models this corresponds to an age of 110,000 years,
an effective temperature of 100,000~K, a luminosity of
1,600~L$_\odot$, and a radius of 0.13~R$_\odot$.  Its mass is 0.53
solar masses.  The faintest white dwarf in our sample has
$\mathrm{F336W}=25.4$, yielding an age 1.2~Gyr, an effective
temperature 8,700~K, a luminosity of $10^{-3}$~L$_\odot$ and a radius
of 0.013~R$_\odot$, one tenth of the radius of the brightest white
dwarf.  Clearly the brightest white dwarf in our sample is not a white
dwarf in the usual sense because thermal energy plays an important
role in the pressure balance of the star.  For this brightest star
$\log g=5.93$ which is less than the minimum of the atmosphere model
grid ($\log g=6$) so we have to extrapolate slightly off of the grid,
but only for this brightest star. For the simulations in
\S\ref{sec:monte-carlo-simul} we did not use this particular model,
but similar ones of the same white dwarf mass with different neutrino
cooling rates or initial metallicities also generated with MESA.

\subsection{Likelihood Function}
\label{sec:likelihood-function}

The model outlined in \S\ref{sec:difuss-lumin-evol} predicts the
number of white dwarfs as a function of magnitude and position.  Let us
divide the space of position and magnitude into bins of width $\Delta R$ and
$\Delta m$ and where the bins are numbered with indices $j$ and $k$
respectively.  The probability of finding $n$ stars in a particular
bin is given by
\begin{equation}
P(n;f(R_j,m_k)) = \frac{\left[f(R_j,m_k) \Delta R \Delta m\right ]^n
  e^{-f(R_j,m_k) \Delta R \Delta m}}{n!}.
\label{eq:37}
\end{equation}
Now we imagine dividing the sample into so many bins that there
is either a single star in a bin or no stars at all, we have
\begin{eqnarray}
P(n;f(R_j,m_k)) &=& e^{-f(R_j,m_k) \Delta R  \Delta m} \times \nonumber \\ 
& &  \left \{
\begin{array}{ll}
f(R_j,m_k) \Delta R \Delta m  & ~~\textrm{one star} \\
1 & ~~\textrm{no star}
\end{array} \right . .
\label{eq:38}
\end{eqnarray}
We can define the likelihood as the logarithm of the product of the
probabilities of observing the number of stars in each bin.  Since 
the bins are so small we can replace $R_j$ and $m_k$ for the bins with
stars in them with the measured values for that particular star $R_i$
and $m_i$.  This gives the so-called ``unbinned likelihood'' 
of observing the sample as follows 
\citep{1979ApJ...228..939C,1996ApJ...461..396M,FermiLikely}
\begin{equation}
\log L = \sum_i \log f(R_i,m_i)  - \sum_{j,k} f(R_j,m_k) \Delta R
\Delta m.
\label{eq:39}
\end{equation}
We have dropped the constant widths of the bins from the first term
which is a sum over the observed stars; consequently, the absolute
value of the likelihood is not important, just differences matter.
The second term is a sum 
over the really narrow (and arbitrary) bins that we have defined, so
we have
\begin{equation}
\sum_{j,k} f(R_j,m_k) \Delta R \Delta m = \int f(R_j,m_k) dR dm = N_\mathrm{pred}
\label{eq:40}
\end{equation}
where $N_\mathrm{pred}$ is the number of stars that the model predicts
that we will observe, so finally we have
\begin{equation}
\log L = \sum_i \log f(R_i,m_i)  - N_\mathrm{pred}
\label{eq:36}
\end{equation}
where the summation is over the observed stars.
The integral for $N_\mathrm{pred}$ when combined with Eq.~\ref{eq:41}
yields
\begin{equation}
N_\mathrm{pred} = {\dot N} \int_0^{t_1} \int_0^{r_\mathrm{max}} \rho(R,t) C(R,m) 
 dR dt
\label{eq:3}
\end{equation}
or
\begin{equation}
N_\mathrm{pred} = {\dot N} \int_{m_0}^{m_1} \int_0^{r_\mathrm{max}} \rho(R,t)
C(R,m) \frac{dt}{dm} 
 dR dm.
\label{eq:4}
\end{equation}
If we take the luminosity function as fixed and try to maximize the
likelihood with respect to the diffusion model
\begin{eqnarray}
\log L &=& \sum_i \log \left [ \dot N \rho(R,t(m_i)) \left . \frac{\partial t}{\partial m}
\right |_i C(R_i,m_i) \right ]  -  N_\mathrm{pred} \nonumber \\ 
 & &
\label{eq:5}
\\
       &=& \sum_i \log \left [ \dot N \rho(R,t(m_i))  \right ] +  \nonumber
\label{eq:6}
\\
& & ~~~ \sum_i \log \left [ \left . \frac{\partial t}{\partial m} \right |_i C(R_i,m_i) \right ]  - N_\mathrm{pred} 
\end{eqnarray}
where the second summation does not depend on the diffusion model so
it is constant with respect to changes in the diffusion model and can be dropped
from the logarithm of the likelihood.  However, it must be included if
one wants to compare different cooling curves, $t(m)$.

\subsection{Magnitude Errors}
\label{sec:magnitude-errors}

An important complication to the analysis is that the measured
magnitudes are not the same as the actual magnitudes of the stars; in
particular the error distribution is not normal or even symmetric.
This transforms the model distribution function via a convolution,
\begin{eqnarray}
f'(R,m) &=& \int_{-\infty}^\infty f(R,m') g(R,m',m-m') dm' \label{eq:24}\\
&=& \int_{-\infty}^\infty f(R,m-\Delta m) g(R,m-\Delta m,\Delta m) d \left (\Delta m\right) \nonumber \\
&=& \int_0^1 f(R,m-\Delta m) dG \label{eq:26}
\end{eqnarray}
where
\begin{equation}
G(R,m,\Delta m) = \int_{-\infty}^{\Delta m} g(R,m-\Delta m,\Delta m') d \left (\Delta m'\right ) \label{eq:27}
\end{equation}
is the cumulative distribution of magnitude errors with the observed
radius and magnitude fixed.  If we calculate the percentiles of the
magnitude error distribution as $\Delta m_j$ we can approximate the
integral as the sum
\begin{equation}
f'(R,m) = \frac{1}{100} \sum_j f(R,m-\Delta m_j),
\end{equation}
so for a given star $i$ we have
\begin{eqnarray}
f'(R_i,m_i) &=& \dot N \sum_j \rho(R,t(m_i-\Delta m_j))  \times \nonumber \\
& & ~~~ t'\left ( m_i - \Delta m_j \right ) C(R_i,m_i-\Delta m_j)
\label{eq:28}
\end{eqnarray}
where $t'=\partial t/\partial m$.  This new function $f'(R_i,m_i)$ can be 
substituted into Eq.~\ref{eq:36} to yield a likelihood including the 
magnitude errors.  We will assume that the magnitude errors do not affect our
estimate of $N_\mathrm{pred}$; this simplifies the analysis.  We will verify 
our technique with Monte Carlo simulations in
\S\ref{sec:monte-carlo-simul}.

\subsection{Constraining the luminosity function}
\label{sec:lf}

We can construct a maximum likelihood estimator of the luminosity
function of the white dwarfs or alternatively the cooling curve as
follows
\begin{eqnarray}
f(R,m) &=& \frac{dN}{dmdR} = {\dot N} \rho(R,t) \frac{\partial t}{\partial m} 
\label{eq:7}
\\
&=& {\dot N} \rho(R,t) \sum_i A_i \delta(m-m_i)
\label{eq:29}
\end{eqnarray}
where $i$ is an index that runs over the observed stars.  With this
model we can define a likelihood function for the stars that we
observe
\begin{equation}
\log L = \sum_i \log \left [ {\dot N} A_i \rho(R_i,t_i) \right ] - N_\mathrm{pred}
\label{eq:8}
\end{equation}
where a multiplicative constant (infinite in this case) and the completeness for each star 
have been dropped from the logarithm.

Substituting the trial luminosity function Eq.~\ref{eq:7} yields
\begin{equation}
N_\mathrm{pred} = \sum_i A_i {\dot N} \int_0^{r_\mathrm{max}} \rho(R,t_i)
C(R,m_i) dR.
\label{eq:9}
\end{equation}
If we maximize the likelihood with respect to the values of $A_k$ we
obtain
\begin{equation}
\frac{\partial \log \left [ \rho(R_i,t_i) C(R_i,m_i)\right ]}{\partial A_k} =
\frac{\partial \log \left [ \rho(R_i,t_i) \right ] }{\partial t_i} \frac{\partial t_i}{\partial
  A_k}
\label{eq:10}
\end{equation}
where 
\begin{equation}
\frac{\partial t_i}{\partial A_k} = \left \{ \begin{array}{lcl}
1 & \textrm{if} & k\leq i \\
0 & \textrm{if} & k> i 
\end{array} \right .
\label{eq:11}
\end{equation}
so
\begin{equation}
\sum_i \frac{\partial \log \left [ A_i \rho(R_i,t_i) C(R_i,m_i)\right ]}{\partial A_k} = \frac{1}{A_k} + 
\sum_{i\geq k} \frac{\partial \log \left [ \rho(R_i,t_i) \right ] }{\partial t_i}
\label{eq:12}
\end{equation}
and 
\begin{equation}
\frac{\partial \log \left [ \rho(R,t) \right ] }{\partial t} =
\frac{R^2}{4 \kappa (t+t_0)^2} - \frac{1}{t+t_0}
\label{eq:13}
\end{equation}
Taking the derivative of $N_\mathrm{pred}$ yields the second part of
the variance in $\log L$,
\begin{eqnarray}
\frac{\partial N_\mathrm{pred}}{\partial A_k} &=&
{\dot N} \int_0^{r_\mathrm{max}} \!\!\!\! \rho(R,t_k)
C(R,m_k) dR + \\ \nonumber 
& & ~~~ 
\sum_i A_i {\dot N} \int_0^{r_\mathrm{max}} \frac{\partial
  \rho(R,t_i)}{\partial t_i} \frac{\partial t_i}{\partial A_k}
C(R,m_i) dR 
\label{eq:14} \\
 &=&
{\dot N} \int_0^{r_\mathrm{max}} \!\!\!\! \rho(R,t_k)
C(R,m_k) dR + \\ \nonumber
& & ~~~
\sum_{i\geq k} A_i {\dot N} \int_0^{r_\mathrm{max}} \frac{\partial
  \rho(R,t_i)}{\partial t_i} C(R,m_i) dR.
\label{eq:15}
\end{eqnarray}
Combining these results with $\partial \log L/\partial A_k=0$ yields an
equation of the form
\begin{eqnarray}
\frac{1}{A_k} &=& 
{\dot N} \int_0^{r_\mathrm{max}} \!\!\!\! \rho(R,t_k)
C(R,m_k) dR  + \\ \nonumber
& & 
\sum_{i\geq k} \Biggl [ A_i {\dot N} \int_0^{r_\mathrm{max}} \frac{\partial
  \rho(R,t_i)}{\partial t_i} C(R,m_i) dR - \\ \nonumber
& & ~~~~ \frac{\partial \log \left [ \rho(R_i,t_i) \right ] }{\partial t_i} \Biggr ] \nonumber
\end{eqnarray}
or a 
matrix equation of the form
\begin{equation}
\sum_i M_{ik} A_i = b_k + \frac{1}{A_k}
\label{eq:16}
\end{equation}
where
\begin{equation}
M_{ik} = \left \{ \begin{array}{lcl}
M_i & \textrm{if} & i\geq k \\
0 & \textrm{if} & i<k 
\end{array} \right .
 \label{eq:17}
\end{equation}
and
\begin{equation}
M_i = {\dot N} \int_0^{r_\mathrm{max}} \frac{\partial
  \rho(R,t_i)}{\partial t_i} C(R,m_i) dR
\label{eq:18}.
\end{equation}
The vector $b_k$ is given by
\begin{equation}
b_k = 
 \sum_{i\geq k} \frac{\partial \log \left [ \rho(R_i,t_i) \right ] }{\partial t_i}
- {\dot N} \int_0^{r_\mathrm{max}} \!\!\!\! \rho(R,t_k)
C(R,m_k) dR
\label{eq:19}
\end{equation}
Although this matrix equation has as many rows as there are stars in
the sample, it is straightforward to solve at least formally in two ways.
The first is
\begin{equation}
A_k = \frac{1}{\sum_i M_{ik} A_i - b_k}
\label{eq:20}
\end{equation}
and the second is
\begin{equation}
A_i = \frac{b_i + (A_i)^{-1}}{M_i} - \frac{b_{i+1}+(A_{i+1})^{-1}}{M_{i+1}}, A_n = \frac{b_n+(A_i)^{-1}}{M_n}.
\label{eq:21}
\end{equation}
The values of $b_i$ and $M_i$, of course depend on the values of $A_i$
through the parameter $t_k$, so the solution must proceed iteratively
perhaps while minimizing with respect to the other parameters of the
model $\kappa$ and $t_0$.  

For each value of the diffusion parameters, we chose to iterate
Eq.~\ref{eq:20} three times to determine the values of $A_k$ within a
loop of two iterations where $M_i$ (Eq.~\ref{eq:18}) and $b_k$
(Eq.~\ref{eq:19}) vary.  Given this new trial luminosity function, the
diffusion parameters are varied to find the maximum likelihood, and
the iterative solution of the luminosity function is repeated.  These
two steps are repeated until the values of the diffusion parameters
from one iteration to the next have changed by less than one part per
hundred.

An interesting limit is when the density distribution is independent
of time.  This understandably yields a simpler solution for $A_i$.  In
particular, $M_i=0$ so
\begin{equation}
A_i = -\frac{1}{b_i} = \left [ {\dot N} \int_0^{r_\mathrm{max}} \!\!\!\! \rho(R,m_i)
C(R,m_i) dR \right ]^{-1}
\label{eq:22}
\end{equation}
where the underlying density distribution is normalized.  The weight
is not the reciprocal of the completeness for star $i$ but rather the
reciprocal of the mean of the completeness of a star with the flux of
star $i$ over the density distribution.  The latter could be evaluated
by taking the mean of the completeness measured for all the stars in
the sample in a magnitude range about star $i$ sufficiently wide to
sample the density distribution.  It is important to note that the
weight is the reciprocal of the mean of the completeness not the mean
of the reciprocal.  If the completeness does not depend strongly on
radius, these two will approximately coincide. Finally if the density
distribution is not known a priori and is not modeled, the weight for
a particular star is simply given by $A_i = \left [ C(R_i,m_i) \right
]^{-1}$.  We call this ``Inv Comp'' in Figs.~\ref{fig:all_diff} 
and~\ref{fig:coolingcurves}.

The likelihood is invariant under changes in the birth rate of the
white dwarfs ($\dot N$) if one also changes the values of $\kappa,
A_i$ and $t_0$ as follows:
\begin{equation}
{\dot N} \rightarrow \alpha {\dot N}, A_i \rightarrow \alpha^{-1} A_i, t_0 \rightarrow \alpha^{-1} t_0 ~\textrm{and}~\kappa \rightarrow \alpha \kappa.
\label{eq:23}
\end{equation}
That is the time scale cannot be fixed without some additional input
such as a theoretical cooling curve or an independent estimate of the
white dwarf birthrate.  The quantities $A_i \dot N$, $\kappa/\dot N$
and $\dot N t_0$ are invariant with respect to this transformation.
In our dataset when we use this modeling technique, we fix the value
of $\dot N$ to the value inferred by the number of giants in our field
as in \citet{coolpap}.

\subsection{Constraining the luminosity function with errors}
\label{sec:lferr}

We start the analysis including magnitude errors with
Eq.~\ref{eq:24} and ~\ref{eq:29} which when combined yield,
\begin{eqnarray}
f'(R,m) &=& \int_{-\infty}^\infty {\dot N} \rho(R,t) C(R,m') \sum_i A_i \delta(m'-m_i) \times \nonumber \\ 
& & ~~~ g(R,m',m-m') dm'\label{eq:25}  \\
&=& {\dot N} \sum_i A_i \rho(R,t_i) g(R,m_i,m-m_i) 
\label{eq:30}
\end{eqnarray}
With this
model we can define a likelihood function for the stars that we
observe
\begin{eqnarray}
\log L &=& \sum_j \log \left [ {\dot N} \sum_i A_i \rho(R_j,t_i) g(R_j,m_i,m_j-m_i) C(R_j,m_i) \right ] 
\nonumber \\
& & ~~~ - N_\mathrm{pred}.
\label{eq:31}
\end{eqnarray}
Note how the magnitude error essentially translates into a spread in
the age of the observed stars.
\begin{eqnarray}
\frac{\partial \log L}{\partial A_k}  &=& \sum_{j,i\geq k} \left \{ \frac{\rho(R_j,t_i) g(R_j,m_i,m_j-m_i) C(R_j,m_i)}{\sum_l A_l \rho(R_j,t_l) g(R_j,m_l,m_j-m_l) C(R_j,m_l)}
\rule{0cm}{0.65cm}\right.
\nonumber \\
& & ~~~  \times \left. \rule{0cm}{0.65cm}
\left [ \delta_{ik}  +  A_i \frac{\partial \log \left [ \rho(R_i,t_i) \right ] }{\partial t_i} \right ] \right \} - \frac{\partial N_\mathrm{pred}}{\partial A_k}.
\label{eq:32}
\end{eqnarray}
Although we have included this additional complication in the
derivations for completeness, we have found that the inclusion of
error convolution in modeling simulated data does not affect the
fitting results, so we did not include this in the modeling of the
dynamics while simultaneously determining the luminosity functions. 

\subsection{Monte Carlo Simulations}
\label{sec:monte-carlo-simul}

To test these techniques in the face of the challenges of
incompleteness and magnitude errors present in our data, we simulated
typically on the order of 10,000 catalogs of the same size as our
dataset with a known luminosity function and a known diffusion model
and attempted to recover the input parameters.  In both cases, the age
of the star is selected first to be between zero and 1.5~Gyr. Given
this age the model cooling curve determines the F336W magnitude.
Second, a radius is selected from the cumulative distribution in
projected radius (Eq.\ref{eq:35}).  Given the radius and magnitude of
the candidate for the catalog, the completeness for this star is
calculated and the star is included in the sample with this
probability.  Finally, the magnitude errors are applied by drawing
from the magnitude error distribution.  We created a sample of 3,167
stars --- the same as in the WFC3 white-dwarf sample.  The fitting
procedure followed two different strategies.

The first was to assume a fixed cooling curve and try to find the
density evolution to determine whether the process is biased in
determining the diffusion parameters and the typical errors.  Finally,
we performed simulations where we did not convolve the models with the
error distribution to calculate the likelihood (in all cases errors
were applied to the simulated data) to see whether the omission of
this step introduced biases.  The second strategy did not assume a
cooling curve and determined the cooling curve as a part of the
process of determining the diffusion.  We did not convolve the cooling
curve with the error distribution while fitting the model; however,
the fake catalogs were created in the same way as in the first
strategy.  In this technique the resulting cooling curve can be
multiplied by a constant factor~(Eq.~\ref{eq:23}), so we determine the
values of $\kappa$ and $t_0$ by fixing the value of $\dot N$ to the
one used to build the catalog. This also fixes the age estimates of
all of the white dwarfs in the sample.
\begin{figure}
\includegraphics[width=\columnwidth]{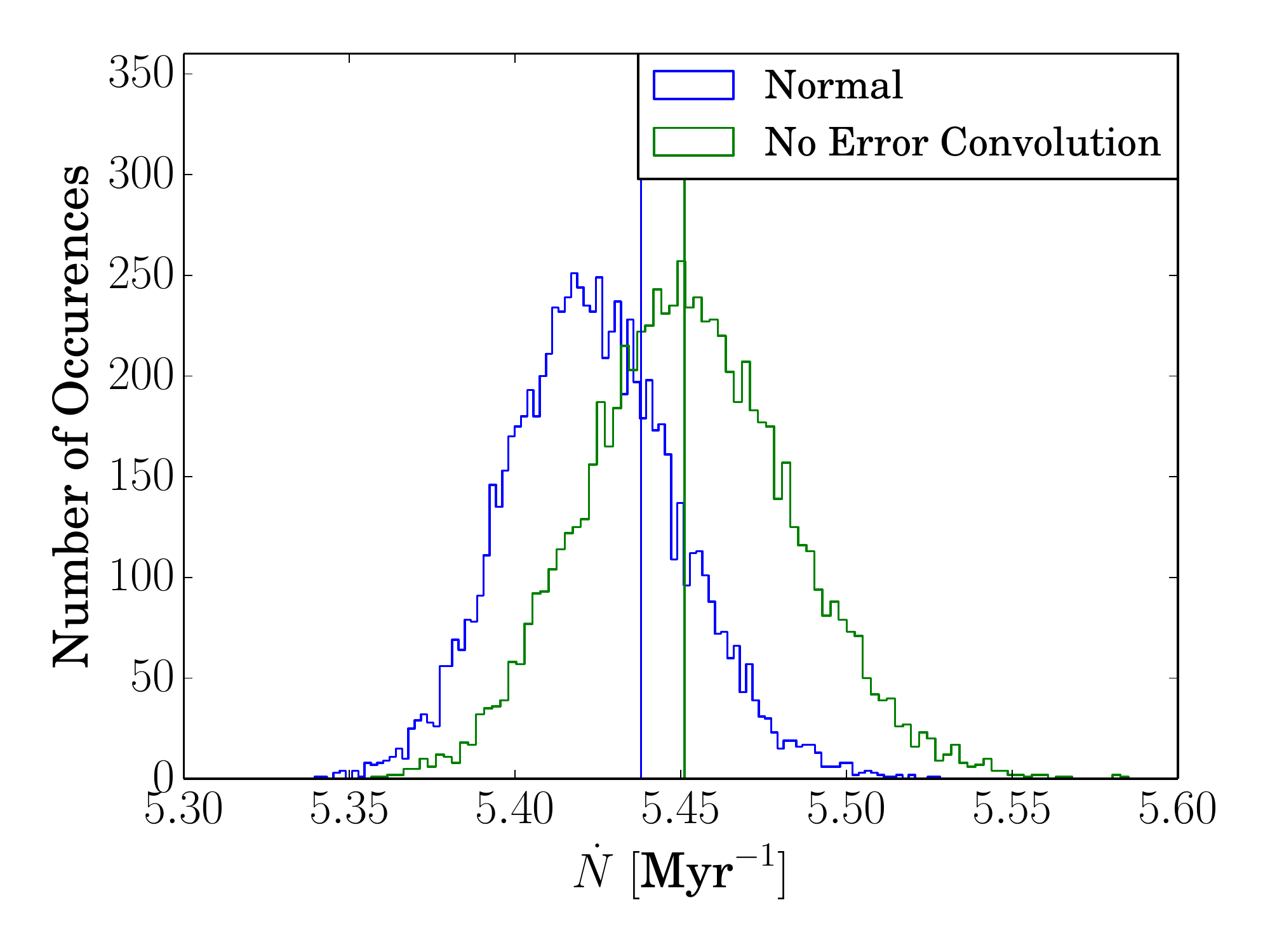}
\includegraphics[width=\columnwidth]{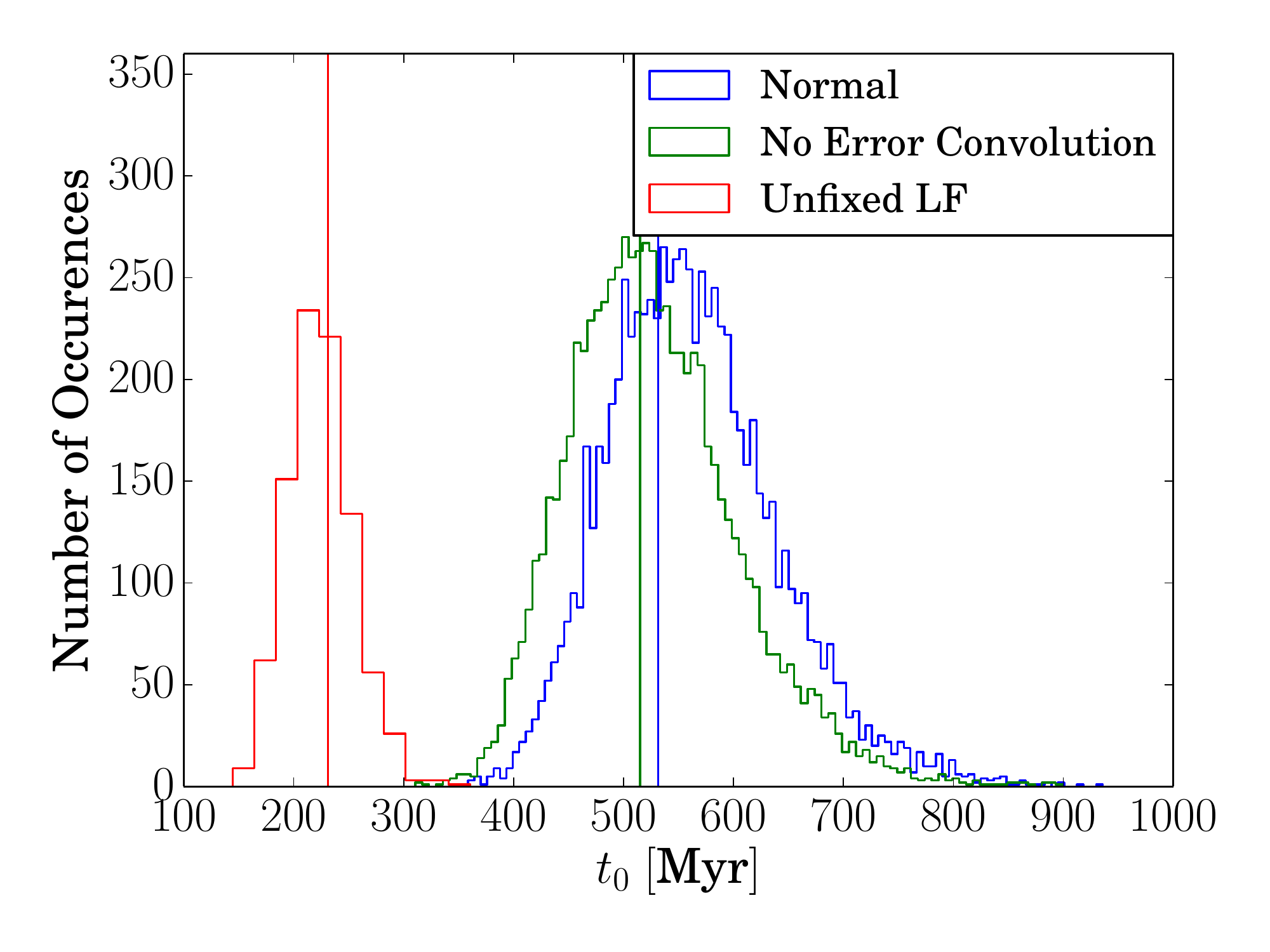}
\caption{{\em Upper panel:} The values of the fitted parameters are
  typically unbiased with respect to the input values in the
  simulations, here $\dot N$ is depicted.  The input values for the
  two types of simulations are given by the vertical lines. 
{\em Lower panel:} The values of $t_0$.}
\label{fig:meNdott0}
\end{figure}

\begin{figure}
\includegraphics[width=\columnwidth]{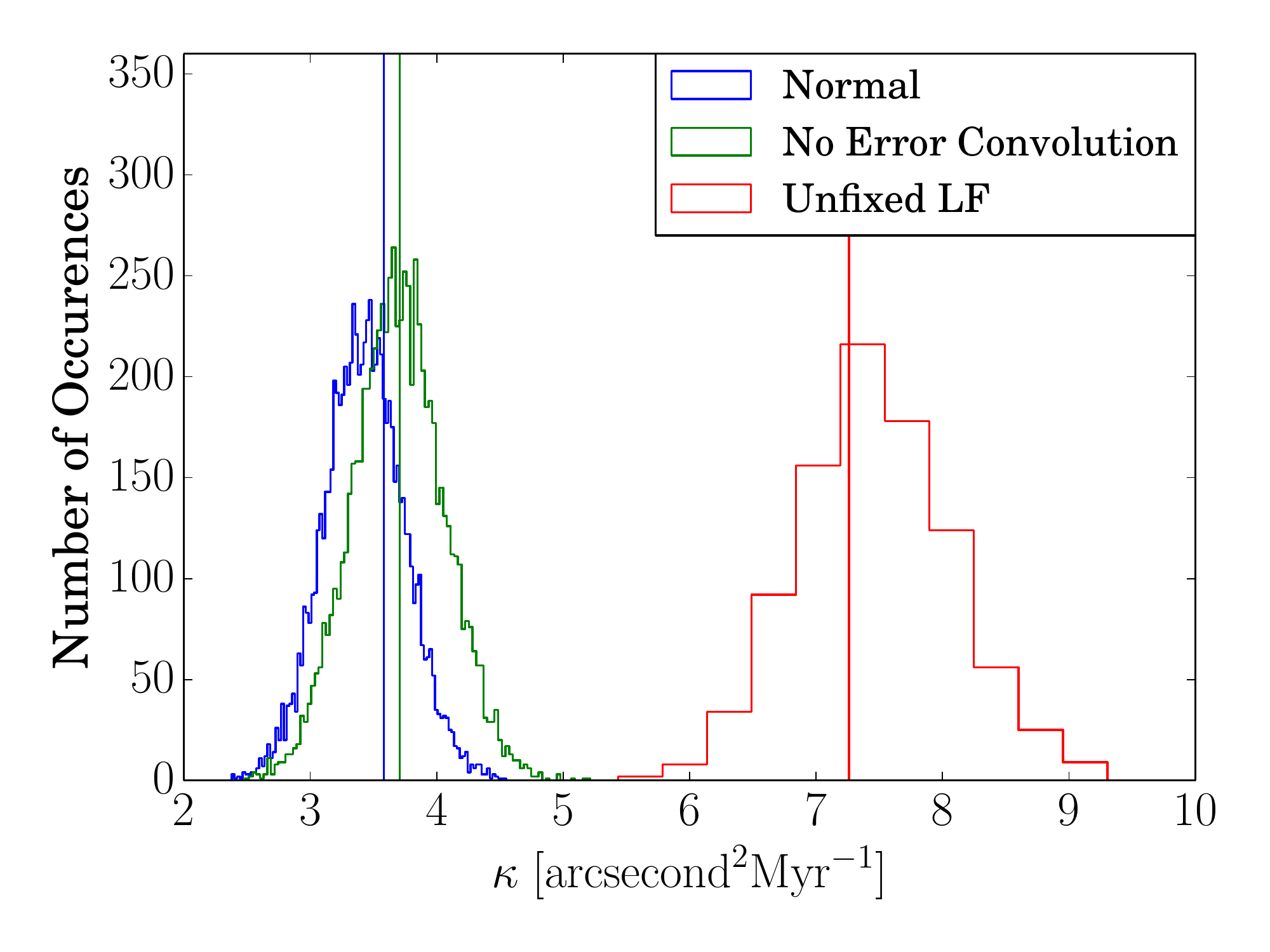}
\includegraphics[width=\columnwidth]{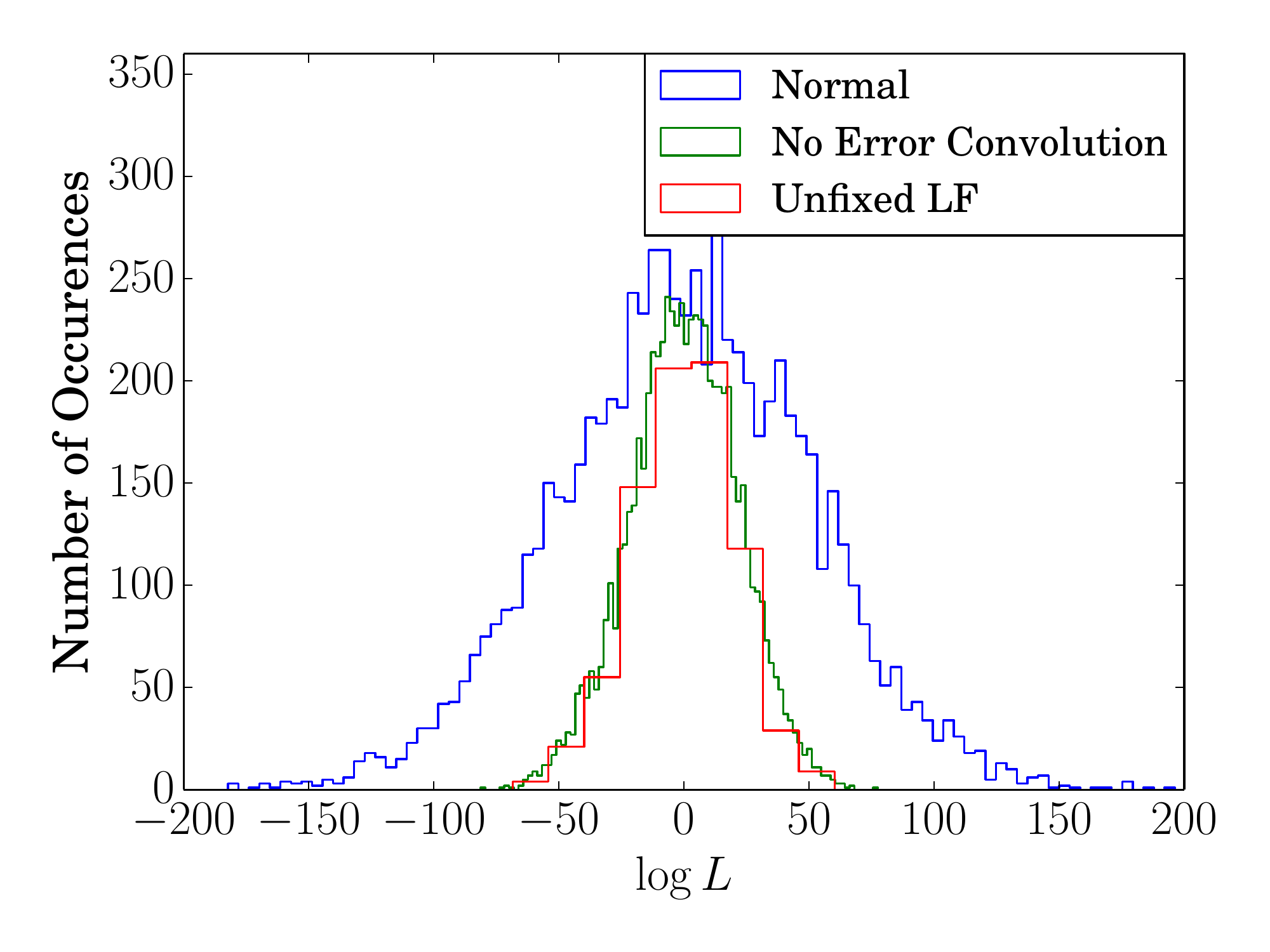}
\caption{{\em Upper panel:} The values of the fitted parameters are
  typically unbiased with respect to the input values in the
  simulations.  Here $\kappa$ is depicted.  The input values for the
  two types of simulations are given by the vertical lines.
 {\em Lower panel:} The distribution of $\log L$ is significantly wider when
  error convolution is included in the fitting process. }
\label{fig:mekappaL}
\end{figure}

\begin{table}
\begin{center}
\caption{Parameters from the Monte Carlo Simulations: Means, Standard Derivations, Input Values}
\label{tab:simulation}
\begin{tabular}{l|cccc}
\hline
Technique            & $\kappa$        & Input  & $t_0$          & Input     \\
\hline
Full Modeling        &  $3.4 \pm 0.3$  & 3.58   &  $560 \pm 80$  & 531  \\
No-Error Convolution &  $3.7 \pm 0.4$  & 3.71   &  $530 \pm 80$  & 515  \\
Unfixed LF           &  $7.5 \pm 0.6$  & 7.26   &  $220 \pm 30$  & 231  \\
\hline
\multicolumn{3}{c}{} \\
\hline
                     & $\dot N$        & Input  & $\log L$ &  \\
\hline
Full Modeling        & $5.42 \pm 0.03$ & 5.44   & $\pm 50$ \\
No-Error Convolution & $5.45 \pm 0.03$ & 5.45   & $\pm 22$ \\
Unfixed LF           & $-$                      & $\pm 20$ \\
\hline
\end{tabular}
\end{center}
\end{table}

The results of these simulations are depicted in
Fig.~\ref{fig:meNdott0}~and~\ref{fig:mekappaL} and in
Tab.~\ref{tab:simulation}.  The key results of the simulations are
that the likelihood fitting of the diffusion model results in an
unbiased estimate of the diffusion parameters regardless of whether
the fitting technique includes the magnitude errors
(\S\ref{sec:magnitude-errors}).  Furthermore, even when one fits for
the luminosity function as well one can obtain reliable estimates of
the diffusion model without prior knowledge of the cooling curve; of
course, in this latter case the timescales of the diffusion rely on an 
independent estimate of the birth rate of the white dwarfs $\dot N$.
Observationally, this is determined from a sample of giant stars
numbering in the thousands (see Fig.~\ref{fig:CMD}) so the statistical
error in this determination is small.  Typically the birth rate is
recovered with an uncertainty of less than one percent and the
diffusion rate with an uncertainty of ten percent and $t_0$ with an
uncertainty of fifteen percent.  The errors in $t_0$ and $\kappa$ are
correlated so the error in $t_0 \kappa$ is typically less than ten
percent.

\begin{figure}
\includegraphics[width=\columnwidth]{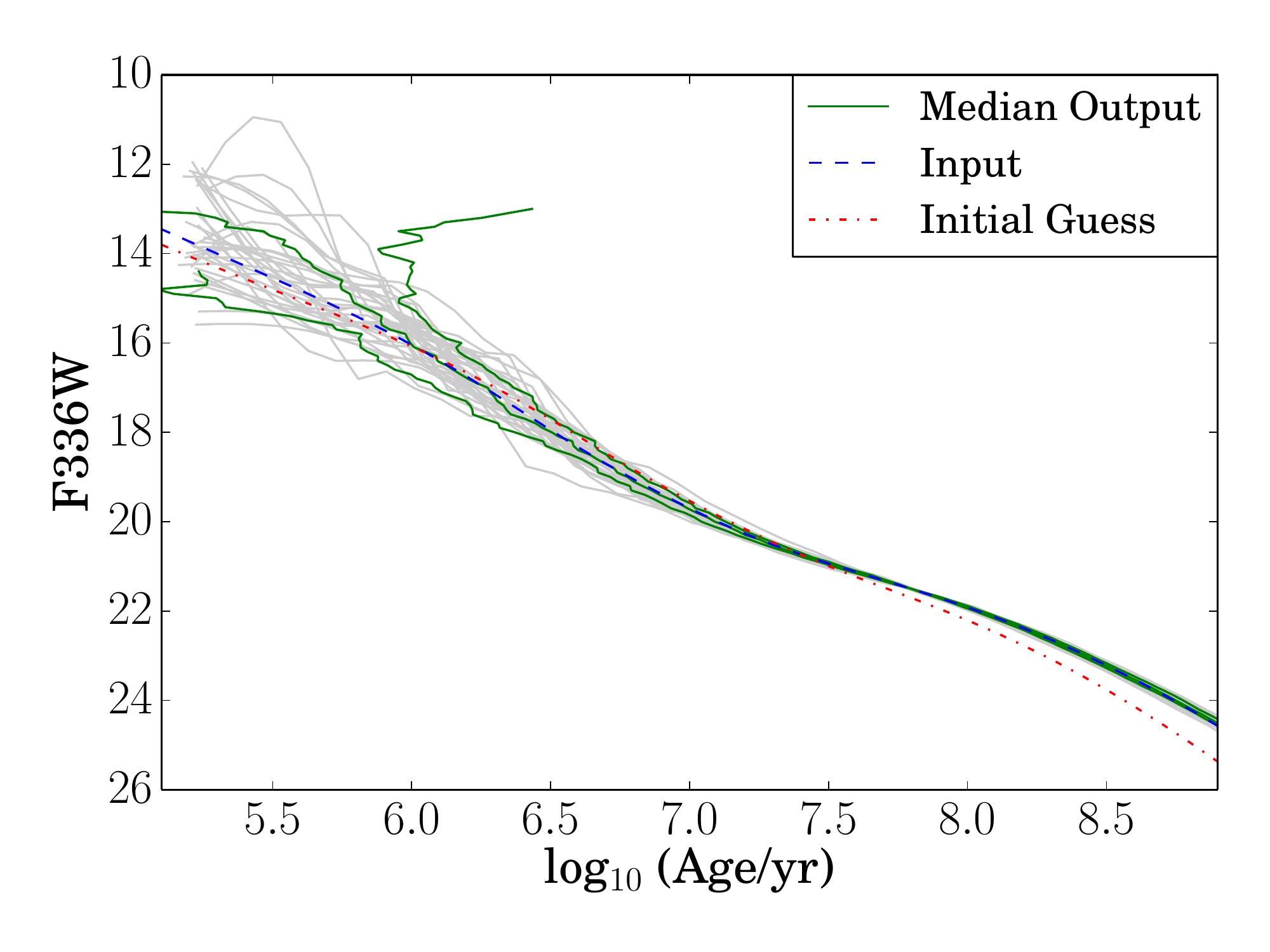}
\includegraphics[width=\columnwidth]{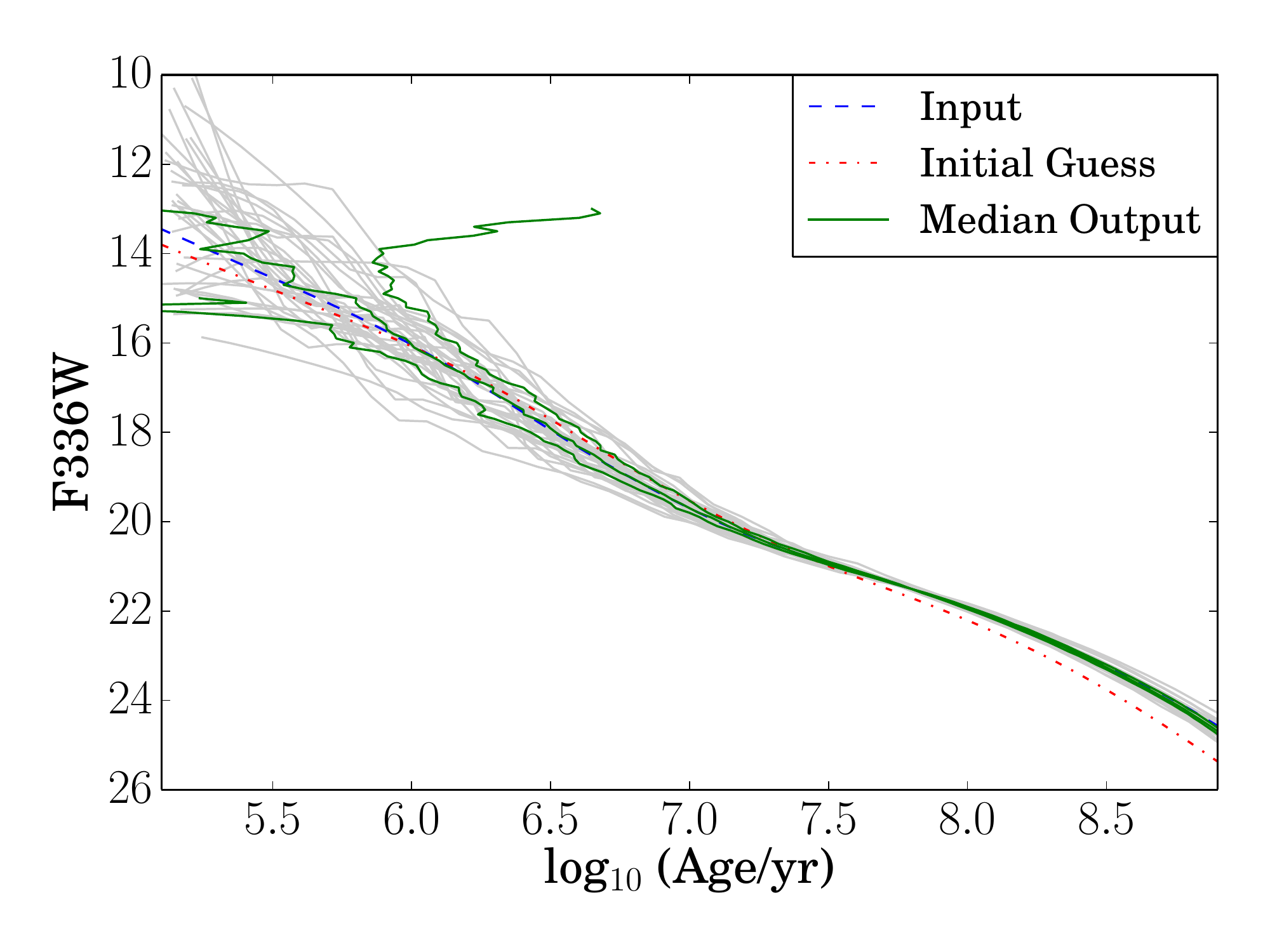}
\caption{{\em Upper panel:} Monte Carlo simulations of an evolving
  density distribution that are fit by an evolving distribution yield
  an unbiased estimate of the cooling curve.  {\em Lower panel:} If
  the evolving density distribution is not included in the fitting the
  resulting cooling curve is typically steeper than the input cooling
  curve; however, this difference is subtle.
  The initial guess is the starting point for the iterative solution
  of the cooling curve.}
\label{fig:all_bunch}
\end{figure}

\begin{figure}
\includegraphics[width=\columnwidth]{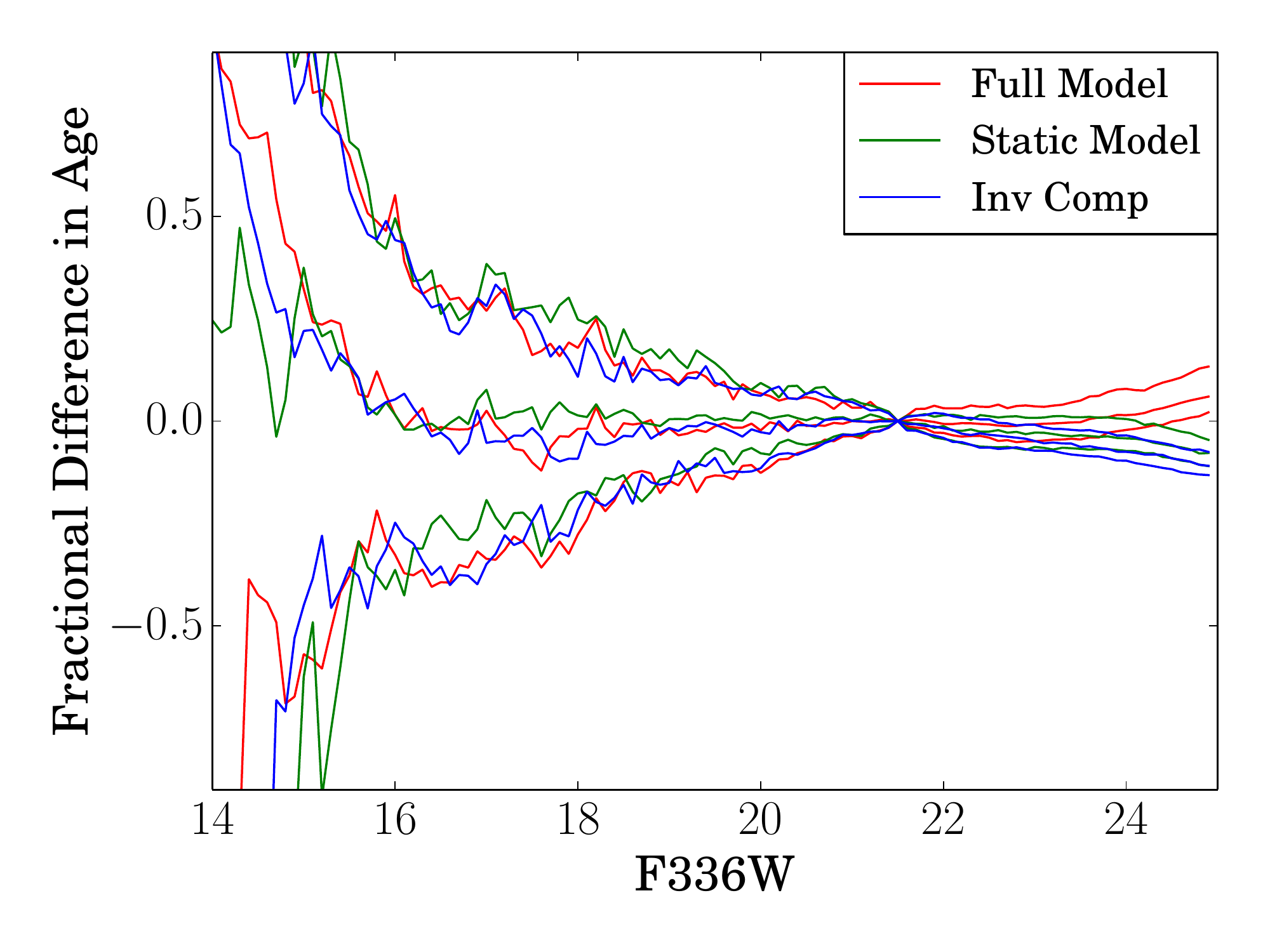}
\caption{The relative differences between the first, second and 
  third quartiles and the input model for the different luminosity 
  function fitting techniques.}
\label{fig:all_diff}
\end{figure}

In the second type of simulation, we found the density evolution along
with an estimate of the cooling curve, so this cooling curve can be
compared with the input cooling curve for the simulations.
Furthermore, the determination of the cooling curve is iterative, so
we have to give an initial guess of the curve.  The input, the initial
guess and the results are given in Fig.~\ref{fig:all_bunch}.  We can
also fit for just the cooling curve and assume that the density
distribution does not evolve or not assume a density model at all and
use the per star completeness as outlined in \S\ref{sec:lf}.   
Fig.~\ref{fig:all_diff} highlights the difference between the model
age and the inferred age with the various likelihood techniques.  For
young white dwarfs the uncertainties are large (because there are few
young white dwarfs in the sample), but for old white dwarfs there is a
small bias of order of ten percent in the inferred age, the sign of
which depends on the technique.  Again this is on the order of the relative
errors in the diffusion parameters.

\section{Results}
\label{sec:results}
\begin{figure*}
\includegraphics[width=0.33\textwidth]{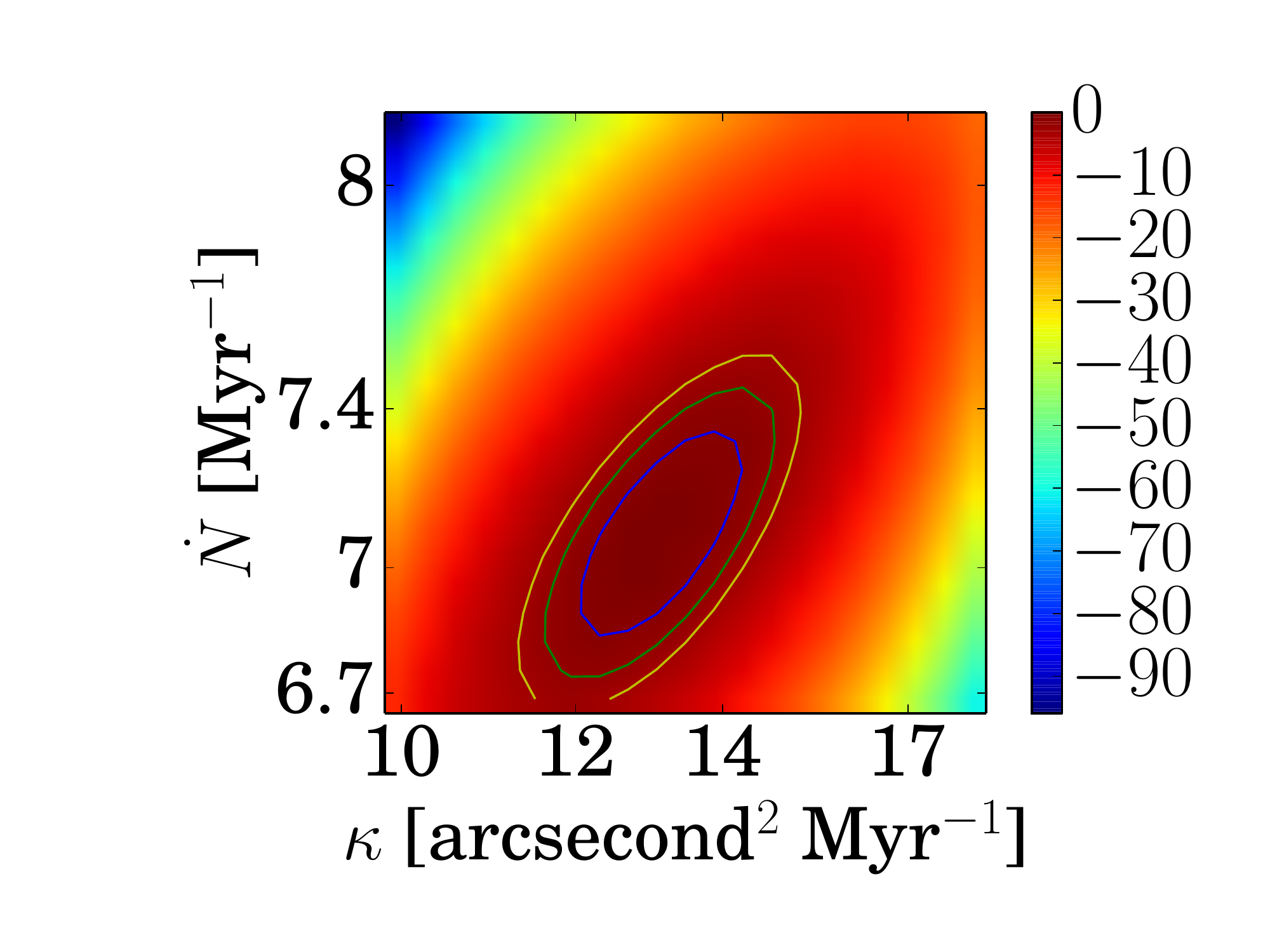}
\includegraphics[width=0.33\textwidth]{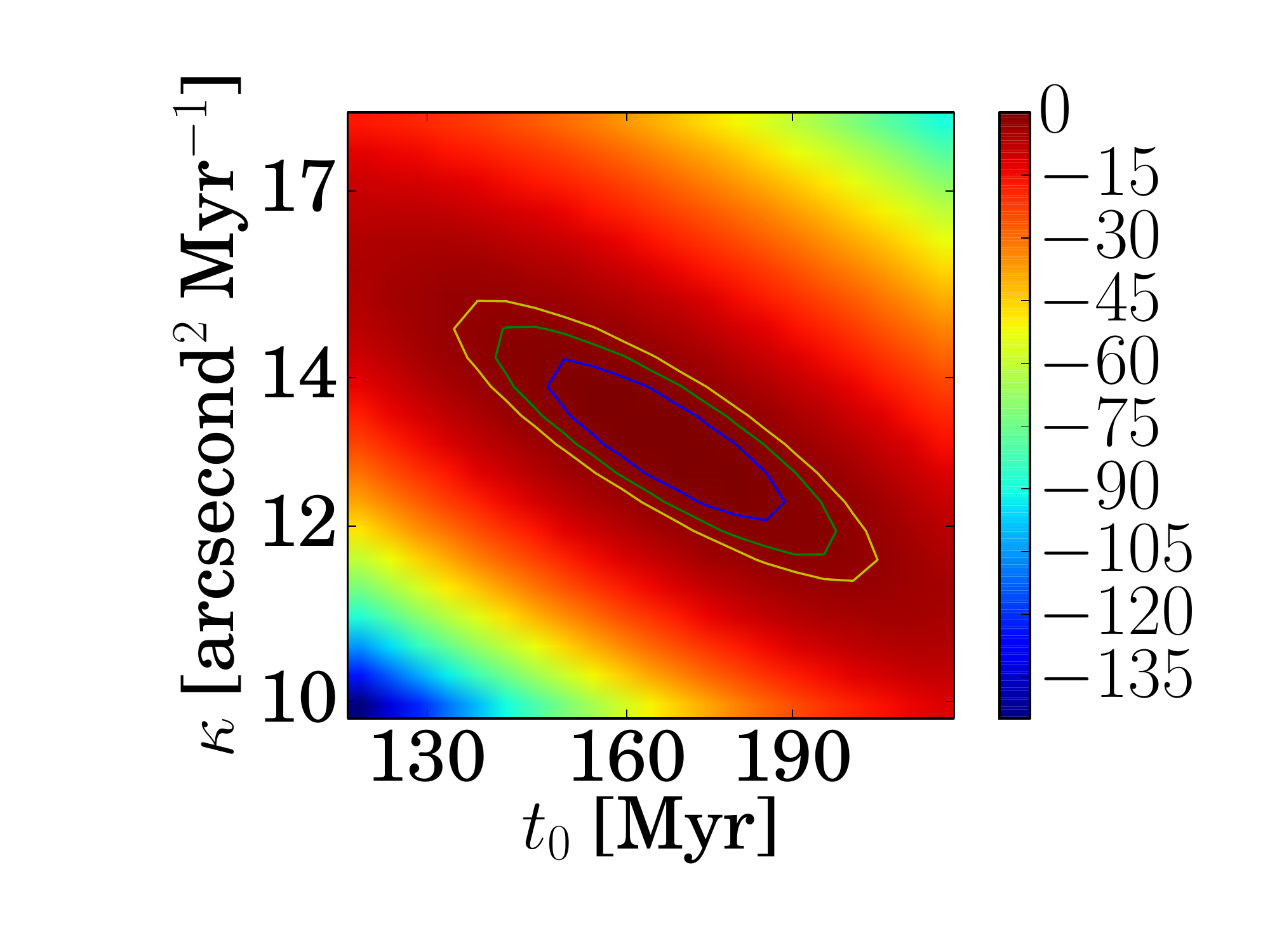}
\includegraphics[width=0.33\textwidth]{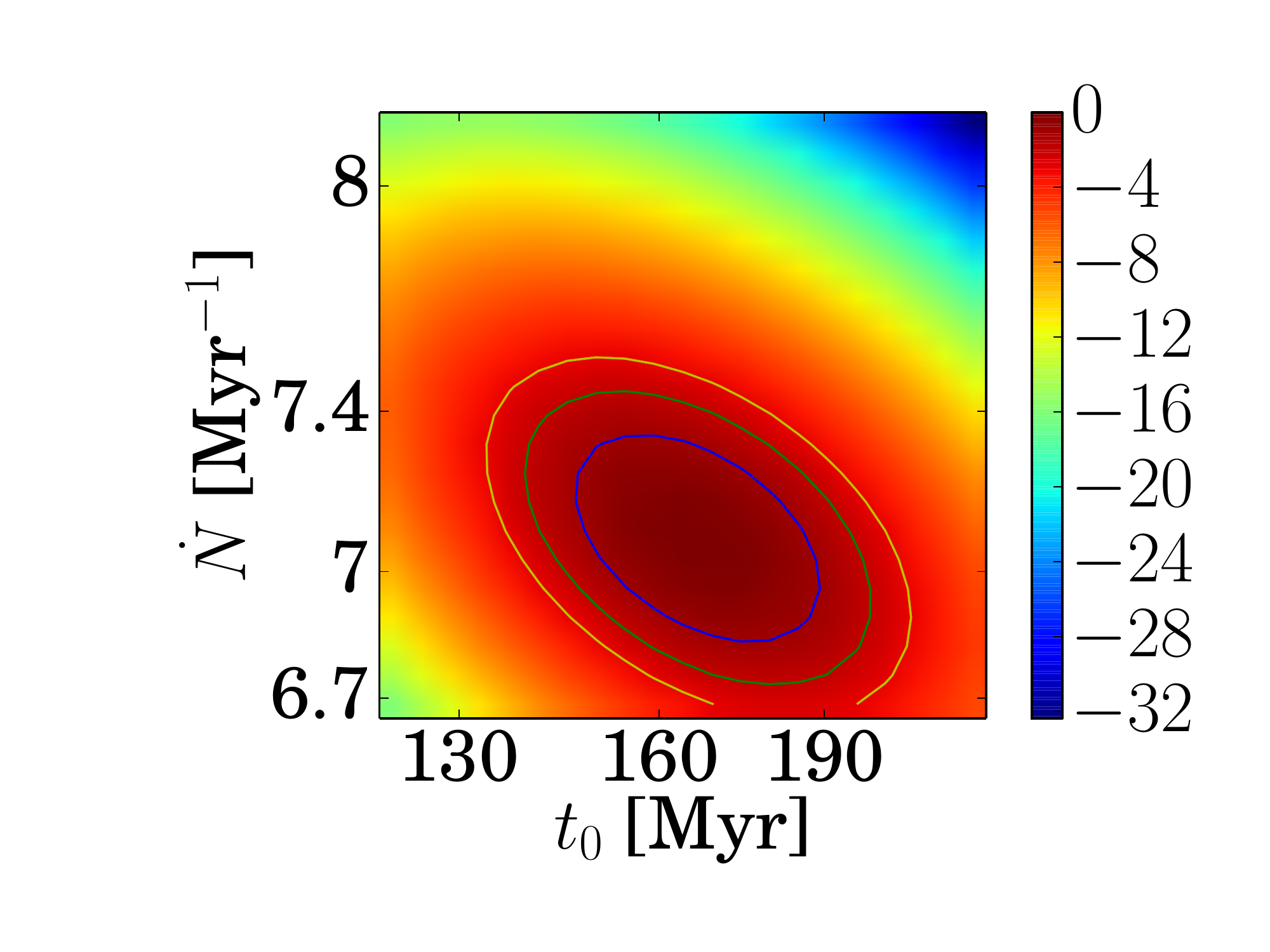}

\includegraphics[width=0.33\textwidth]{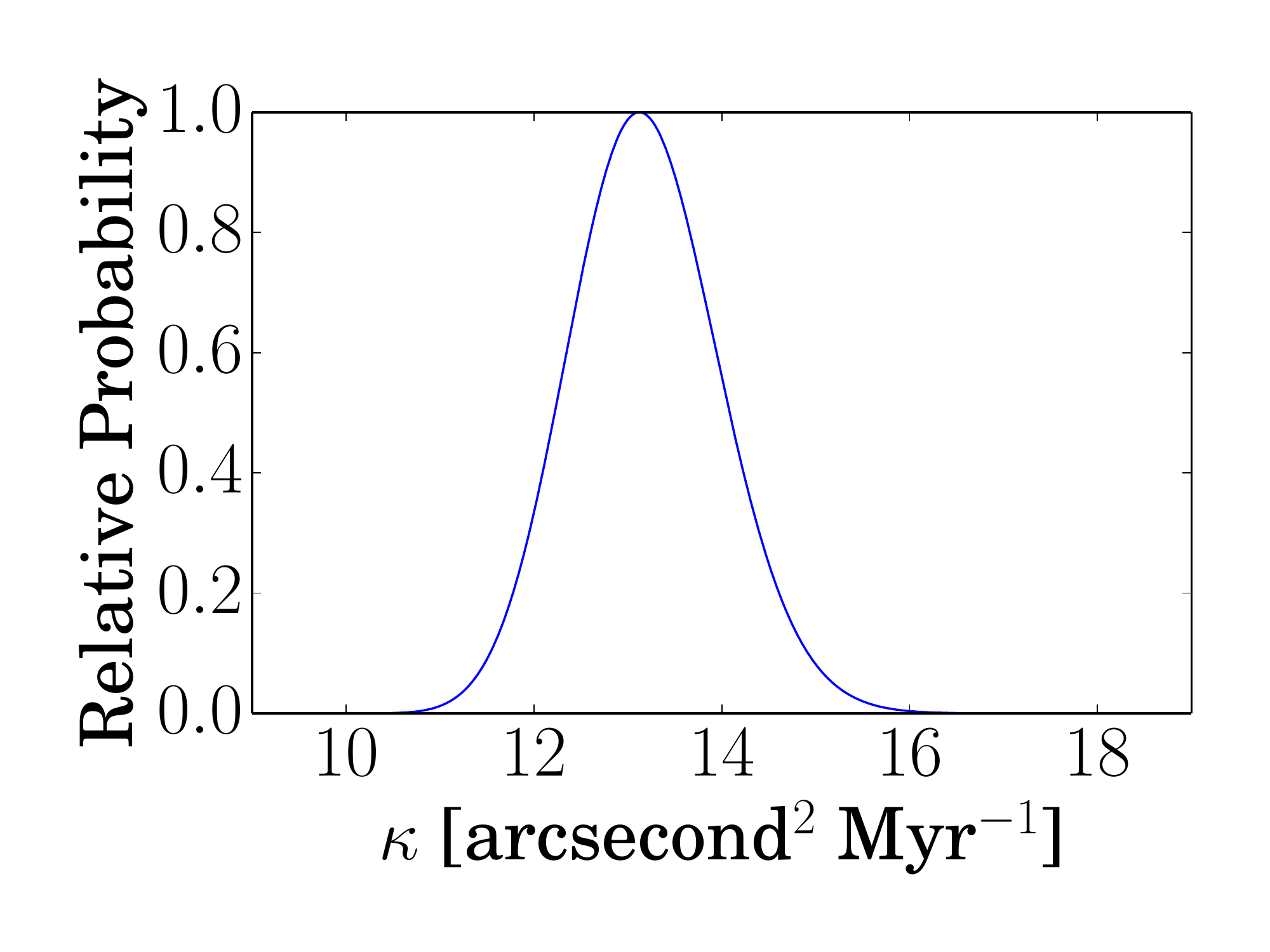}
\includegraphics[width=0.33\textwidth]{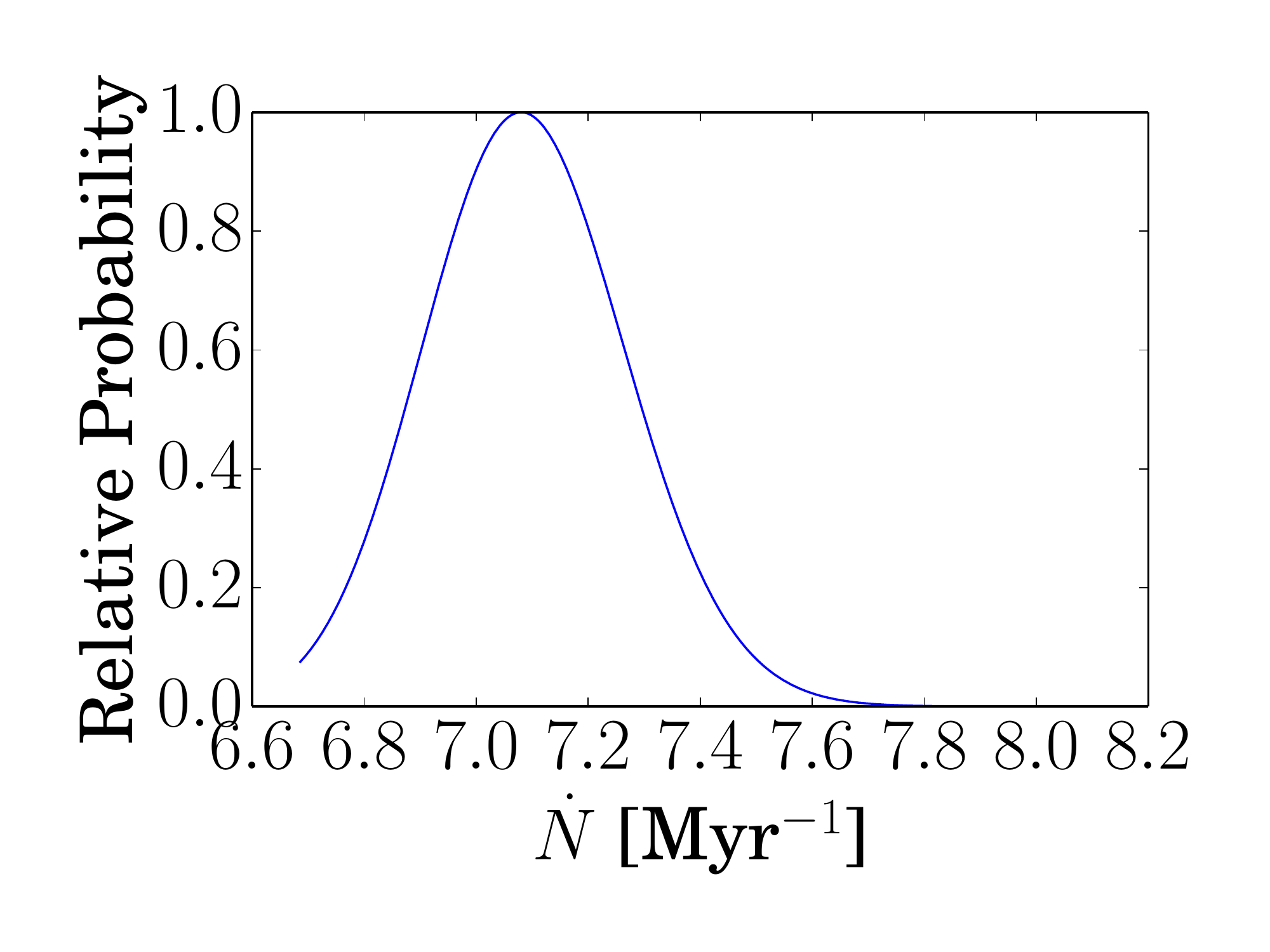}
\includegraphics[width=0.33\textwidth]{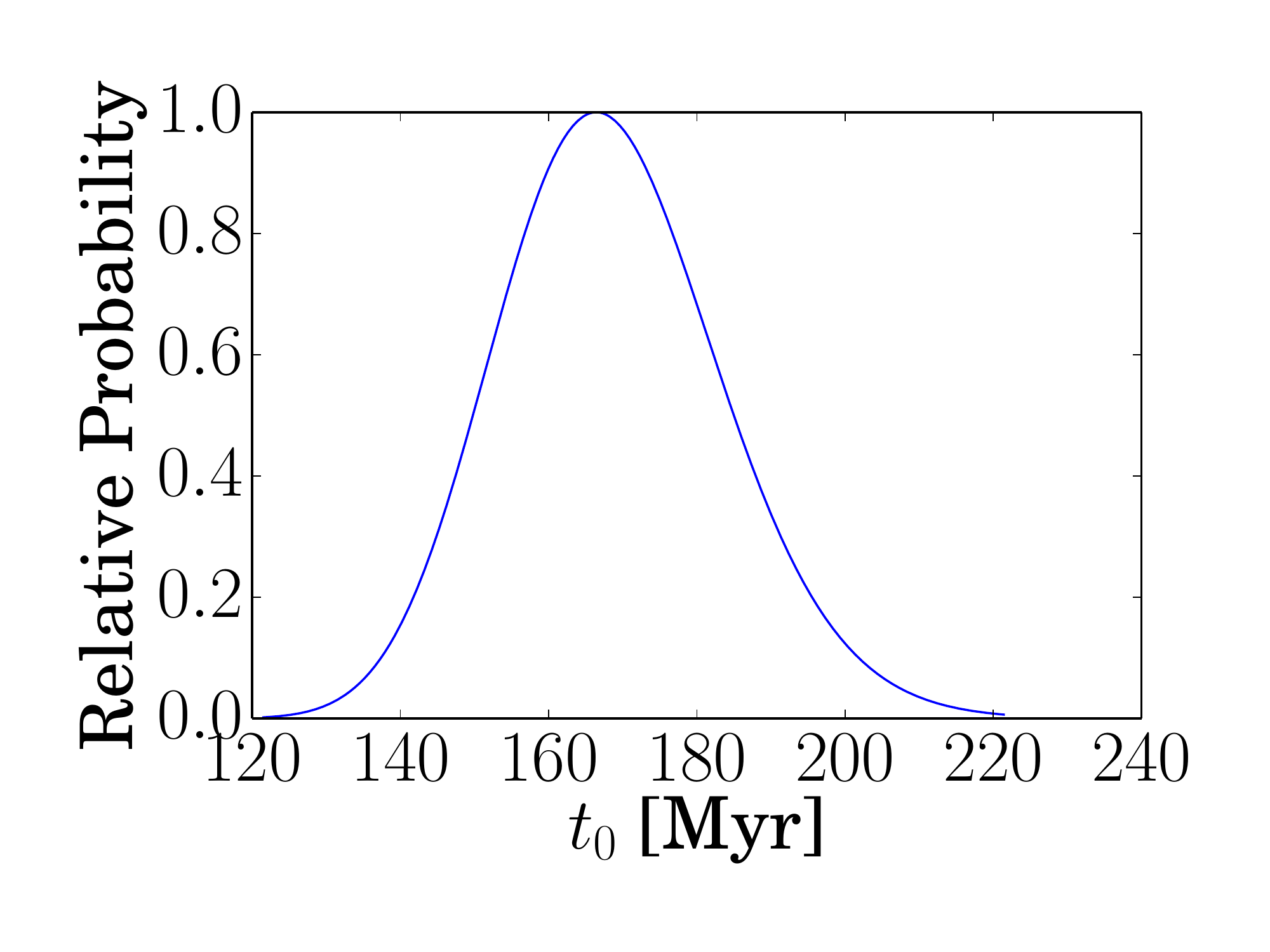}
\caption{The posterior probabilities of the best-fitting parameters.
  The upper panels depict the covariance among the various parameters.
  The contours trace probabilities a factor of $e$, $e^2$ and $e^3$
  smaller than the maximum likelihood.  The color scale also gives the
  natural logarithm of the relative likelihood.  The lower panels give
  the likelihood as a function of a single parameter with the other
  parameters integrated out. }
\label{fig:bestfit}
\end{figure*}

The results of the diffusion model fitting are given in
Tab.~\ref{tab:diffparam}.  The results do not depend strongly on the
modeling technique, especially the assumed cooling curve for the
white dwarfs.  The inferred relaxation times are also in good
agreement with the value tabulated by \citet{1996AJ....112.1487H}.
Fig.~\ref{fig:bestfit} shows the posterior probability distribution
for the various parameters and how the uncertainties are correlated
with each other.  An important conclusion is that the no-diffusion
model ({\em i.e.} $\kappa=0$) is excluded at high confidence.

\begin{table}
\begin{center}
\caption{Diffusion parameters from the likelihood fitting. The values
  of $\sigma_0$ and $t_r$ are given by $\sqrt{2\kappa t_0}$ and
  $r_c^2/\kappa$ where $r_c = 22\arcsec$. The posterior probabilities for
  the full model are given in Fig.~\ref{fig:bestfit}.}
\label{tab:diffparam}
\begin{tabular}{l|rrrrrr}
\hline
      & \multicolumn{1}{c}{$\kappa$} &
 \multicolumn{1}{c}{$t_0$} & 
\multicolumn{1}{c}{$\sigma_0$} & 
\multicolumn{1}{c}{$\dot N$} & 
$t_r(r_c)$ \\ 
Model & \multicolumn{1}{c}{[$(\arcsec)^2$ Myr$^{-1}$]} & \multicolumn{1}{c}{[Myr]} & \multicolumn{1}{c}{[$\arcsec$]} & \multicolumn{1}{c}{[Myr$^{-1}$]} & \multicolumn{1}{c}{[Myr]} \\ \hline
Full           &  13.1~~~ &  166~ & 66~   & 7.07~~  & 37~~ \\
No Errors       &  13.1~~~ &  166~ & 66~   & 7.07~~  & 37~~ \\
2 Gaussians    & 12.8~~~  &  14.9~ & 19.5$~\!$   & 1.90~~  & 38~~ \\
~~(No Errors)   &          &  260~ & 82~   & 5.32~~  & \\
Free LF        &  9.80~~~ &  241~ & 69~   & 8.10~~  & 49~~ \\ \hline
\end{tabular}
\end{center}
\end{table}

To find whether we could better fit the radial distribution with a sum
of Gaussians, we performed the fitting with two and three
Gaussians. We did not include the error convolution in the fitting
models. The fit with two Gaussians has a value of $\log L$ that is
lower by 75 from a fit with a single Gaussian.  From
Fig~\ref{fig:mekappaL} we can see that this is a significantly better
fit.  However, the decrease in $\log L$ by adding a third Gaussian is
only 2; furthermore, the third Gaussian has a very low value of $\dot N$ so it
does not affect the resulting distributions strongly.
Tab.~\ref{tab:diffparam} shows that the diffusion parameters from the
two-Gaussian fit only differ slightly from the one-Gaussian fits.
In any case these differences lie within the statistical errors.  We
can compare the best-fitting model density distributions as a function
of time with the observed (completeness corrected) density
distributions for several age ranges of white dwarf.  The diffusion
model for the median age of the white dwarfs in each bin is depicted
with a solid line for the one-Gaussian model and a dot-dashed line for
the two-Gaussian model.  The two-Gaussian model does a better job at
following the distribution of the white dwarfs especially at smaller radii.
\begin{figure}
\includegraphics[width=\columnwidth]{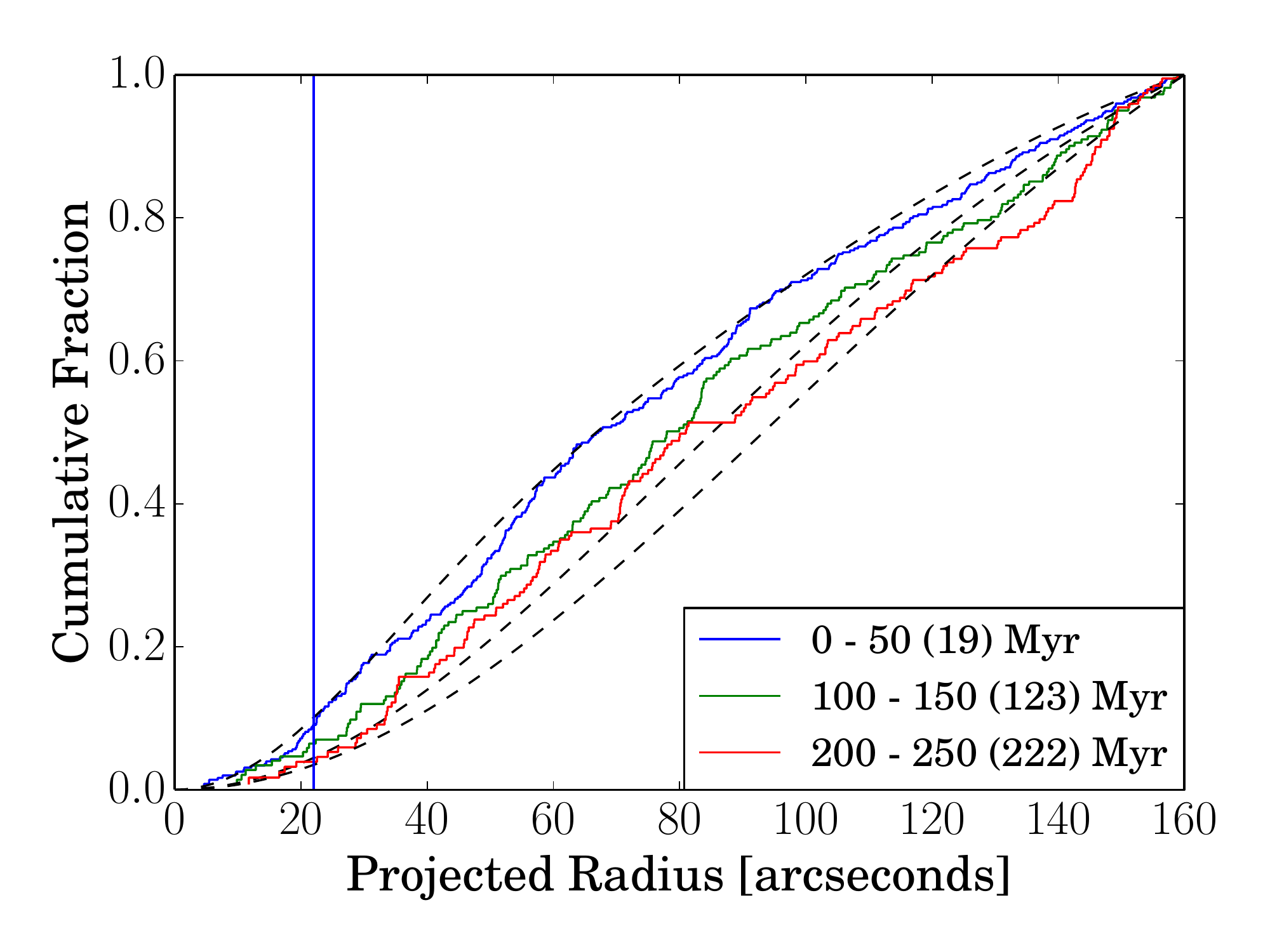}
\caption{Radial distribution of white-dwarf samples of various age
  ranges and median ages (given in parenthesis) and the best-fitting two-Gaussian
  diffusion models superimposed. The core radius from 
\citet{1996AJ....112.1487H} is depicted by the vertical line.}
\label{fig:raddist}
\end{figure}

\subsection{Two-Body Relaxation}
\label{sec:two-body-relaxation}

Fig.~\ref{fig:raddist} depicts the radial distribution of white dwarfs
of various ages.  Each bin is 50~Myr wide, and the bins are centered
on 25~Myr, 125~Myr and 225~Myr.  The evolution at up to a few core
radii (about 60~arcseconds) is dramatic from 25 to 125~Myr and modest
thereafter.  Outside 60~arcseconds the cumulative distributions are
nearly parallel indicating little evolution in this region at early
times.  The simple diffusion models used here assume that the
diffusion coefficient is constant in space and in time, so the models
continue to evolve at late time and for all radii.  At the smaller
radii the white dwarfs reach the distribution corresponding to their
masses after about 100~Myr and stop diffusing.

\begin{figure}
\includegraphics[width=\columnwidth]{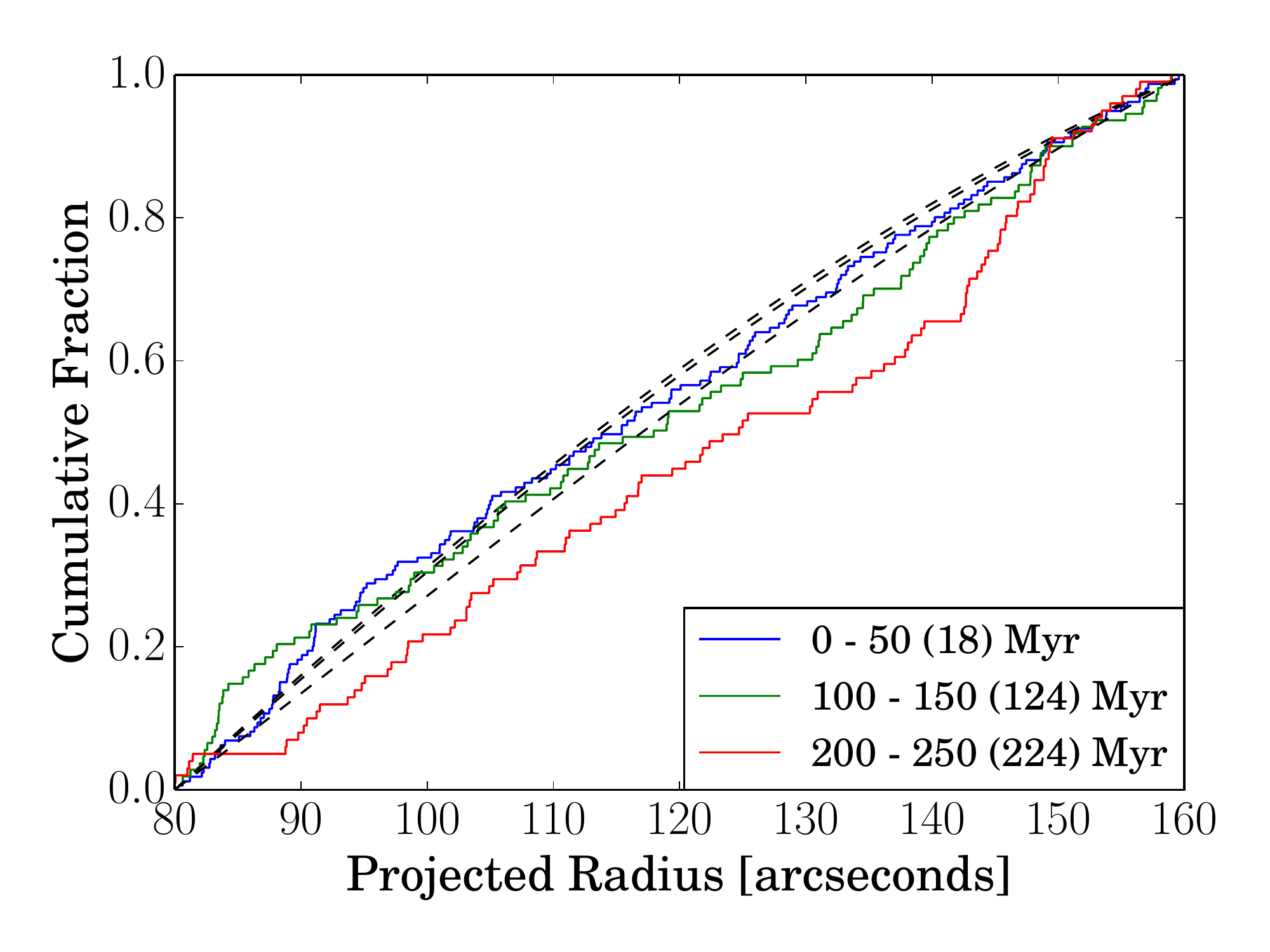}
\caption{Radial distribution of white-dwarf samples of various age
  ranges and median ages (given in parenthesis) and the best-fitting two-Gaussian
  diffusion models superimposed for the outer half of the region.}
\label{fig:raddist2}
\end{figure}

Fig.~\ref{fig:raddist2} focuses on the outer half of the WFC3 field.
Here we see more evolution between the second and third epochs with
little early evolution.  This indicates the increase in the relaxation
time as the stellar density decreases.  The white dwarfs diffuse
modestly over the first 100~Myr and more dramatically during the
second 200~Myr.  The white dwarfs as expected from theoretical
considerations suffer diffusion that is a function of radius and time
and beyond the scope of the simple model used to quantify the
diffusion in this paper.  However, this model does capture the diffusion 
within a few core radii for a few core relaxation times.

\section{Conclusions}
\label{sec:conclusions}

\subsection{Further analysis}
\label{sec:further-analysis}

In this paper we used the Green's function (Eq.~\ref{eq:34}) to model
the diffusion of the stars through the cluster.  We simply took the
initial conditions to be a Gaussian or a sum of Gaussians centered on the
center of the cluster.  This allowed for a simple closed-form
expression for the density function in spherical coordinates and in
projection as well.  Without relaxing the spherical symmetry one could
imagine much more general initial conditions. In fact we have an
estimate of the initial conditions in the form of the projected radial
distribution of the stars on the upper main sequence.  This
distribution could be possibly deprojected as a lowered-isothermal
distribution in phase space \citep{1963MNRAS.125..127M,
  1966AJ.....71...64K} and convolved with the Gaussian Green's
function, Eq.~\ref{eq:34}, to give the expected density distribution
as a function of time.   This technique shares the advantage of the
technique used in this paper that the density distribution can be
guaranteed to be positive because the convolution of the positive
kernel with a positive distribution is necessarily positive; however,
the density distribution even in spherical coordinates is not
available in closed form.

A second strategy would be to expand the initial density distribution
in terms of spherical Bessel functions and spherical harmonics.  If we 
restrict ourselves to an initially spherical distribution we have
\begin{equation}
\rho(r,t) = \int_0^\infty d k a(k) e^{-\kappa k^2 t}
\frac{\sin(kr)}{kr}
\label{eq:44}
\end{equation}
where the coefficients $a(k)$ are determined from the initial density distribution
\begin{equation}
a(k) = \frac{2}{\pi} \int_0^\infty dr \left (k r \right)^2 \rho(r,0) \frac{\sin(kr)}{kr}.
\label{eq:45}
\end{equation}
If the initial density distribution can be well represented with a few
values of $k$, then the density evolution is straightforward to evolve
forward and backward in time; however, it is no longer guaranteed to
be positive even at the initial time if only a range of values of $k$
are considered in $a(k)$.  

\begin{figure}
\includegraphics[width=\columnwidth]{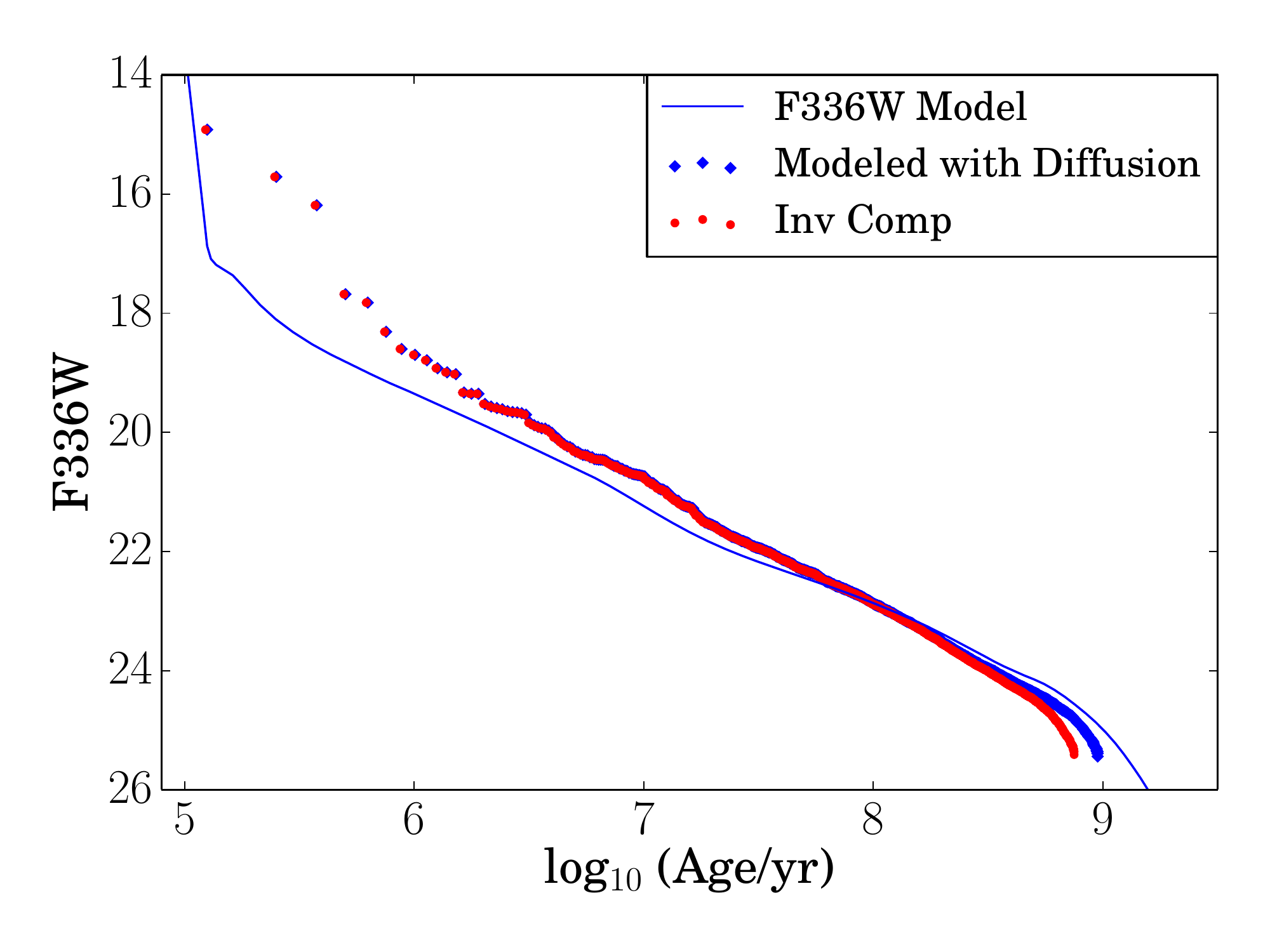}
\caption{Flux in the WFC3 band F336W as a function of time since the
  peak luminosity of the star that we define to be the birth of the
  `white dwarf.  The model curve assumes a true distance modulus of 13.23
  \citep{2010AJ....139..329T} and a reddening of $E(B-V)=0.04$
  \citep{2007A&A...476..243S}.  The ``Inv Comp'' technique ignores
  the effects of diffusion in modeling the stars and uses the completeness
  rate corresponding to the observed magnitude and radius of each star.
}
\label{fig:coolingcurves}
\end{figure}

From the point of view of the likelihood analysis, a natural next step
would be to use the additional information available with the current
observations, {\em i.e.} the flux in the F225W band.  This would
provide an additional constraint on the ages of the white dwarf stars
or alternatively constrain the cooling curve in both bands.  In the
first case one would perhaps get better constraints on the dynamical
evolution and could also fit for the distance and reddening to the
cluster and possibly the mass of the white dwarfs or specifics of the
cooling mechanism.  In the second case one would get a cooling curve
in a second band.  It is straightforward to see that the weights for
the cooling curve in F225W would be the same as in F336W, so simply
plotting the inferred ages of the white dwarfs from
Fig.~\ref{fig:coolingcurves} against the F225W magnitude would yield
the cooling curve in F225W.  The agreement with the F336W model is
poorer at early times but improves with age and lasts until nearly
1~Gyr.   In the context of this paper,
we obtain similar diffusion parameters whether we fit a luminosity
function or assume a theoretical model.

\subsection{Theoretical directions}
\label{sec:theor-direct}

As argued in \S\ref{sec:difuss-lumin-evol} the interactions with other
stars cause the white dwarfs to diffuse in velocity not in
radius.  However, we argued using the virial theorem that this
diffusion in velocity would be manifest as a diffusion in radius as
well. Furthermore, our simple model assumes that the diffusion
coefficient is constant in space and time when in fact with time the
white dwarf distribution approaches that of stars of similar mass so
the diffusion must cease and also at larger radii the diffusion must
happen more slowly.  We see both of these effects in
Fig.~\ref{fig:raddist} and~\ref{fig:raddist2}.  How this diffusion
actually manifests itself could be simulated in two possible ways.

The first is direct numerical simulation of on order of one million
stars that form the central regions of the globular cluster 47 Tucanae.
Although on the face of it, this appears to be a Herculean labor when
the state of the art direct calculation of the two body interactions
in a globular cluster involve merely $\sim 10^5$ stars and the
simulation in question would normally take 100 times longer.
However, we are only interested in the dynamical evolution of the
young white dwarfs over about one hundredth of the age of the cluster
(100-150~Myr out of 10~Gyr).  Secondly, because we are not interested
in the long term evolution of the cluster, neither stellar evolution
nor the dynamics of binaries should play an important role in this
process.  These two simplifications result in a factor of a thousand
speed up to obtain results and these calculations are already
underway.  

The second direction would be to model the diffusion in phase space
using a Fokker-Planck or Monte Carlo scheme
\citep{2011MNRAS.410.2698G,2013ApJS..204...15P,2013MNRAS.430.2960H}.
Such simulations would be more rapid than a direct n-body simulations
and possibly yield more physical insights.

\subsection{Further observations}
\label{sec:further-observations}

Following the arguments of the preceding subsection
\S\ref{sec:theor-direct} a natural direction would be to measure the
proper motions of the white dwarfs in the core of 47 Tuc with a second
epoch of observations.  Because we already have the colors of the
white dwarfs, only observations in a single band would be required and
possibly not as deep as the present set of observations because the
stars have already been detected.  To obtain the most precise
positions and to minimize the crowding, the bluest band would be best,
{\em i.e.} F225W, and possibly over only a portion of the field of the
current data, because here the goal would be to verify the current
result by finding the corresponding signal in velocity space, so a
full sample of 3,000 plus white dwarfs may not be required.  

\subsection{Final remarks}
\label{sec:final-remarks}

We have measured directly for the first time the dynamical relaxation
of stars in a globular cluster. To do this we have introduced new
statistical techniques for the characterization of stellar
populations. These techniques can robustly and straightforwardly
account for high incompleteness and non-Gaussian magnitude errors.
They can be applied to a wide variety of questions from globular
cluster dynamics to galaxy luminosity functions.  There are many
avenues for further investigation such as a more thorough analysis of
the existing data using the information from the second band, the
simulation of the relaxation of young white dwarfs in numerical models
and measuring the proper motions of the young white dwarfs to search
for signatures of relaxation in their velocities as well.

This research is based on NASA/ESA Hubble Space Telescope observations
obtained at the Space Telescope Science Institute, which is operated
by the Association of Universities for Research in Astronomy
Inc. under NASA contract NAS5-26555. These observations are associated
with proposal GO-12971 (PI: Richer).  This work was supported by
NASA/HST grant GO-12971, the Natural Sciences and Engineering
Research Council of Canada, the Canadian Foundation for Innovation,
the British Columbia Knowledge Development Fund. This project was
supported by the National Science Foundation (NSF) through grant
AST-1211719. It has made used of the NASA ADS and arXiv.org.

\bibliography{evolution,mine,globular}
\bibliographystyle{apj}

\end{document}